%% file: main.tex
\documentclass[10pt,conference]{IEEEtran}
\IEEEoverridecommandlockouts

\usepackage[utf8]{inputenc}
\usepackage{xspace}
\usepackage{cite}
\usepackage{graphicx}
\usepackage{url}
\usepackage{booktabs}
\usepackage[flushleft]{threeparttable}
\usepackage{pifont}

\input{micro.tex}

\begin{document}
\title{Are LLMs Correctly Integrated into \\ Software Systems?}

\author{
\IEEEauthorblockN{Yuchen Shao\IEEEauthorrefmark{4}, Yuheng Huang\IEEEauthorrefmark{2}, Jiawei Shen\IEEEauthorrefmark{4}, Lei Ma\IEEEauthorrefmark{2}\IEEEauthorrefmark{3}, Ting Su\IEEEauthorrefmark{4}, Chengcheng Wan*\IEEEauthorrefmark{4}\thanks{* Chengcheng Wan is the corresponding author.}}
\IEEEauthorblockA{
    \IEEEauthorrefmark{4}
    East China Normal University, Shanghai, China
    \IEEEauthorrefmark{2}
    The University of Tokyo, Tokyo, Japan\\
    \IEEEauthorrefmark{3}
    University of Alberta, Edmonton, AB, Canada\\
ycshao@stu.ecnu.edu.cn, yuhenghuang42@g.ecc.u-tokyo.ac.jp, javishen@stu.ecnu.edu.cn,\\ ma.lei@acm.org, tsu@sei.ecnu.edu.cn, ccwan@sei.ecnu.edu.cn}
}

\maketitle
\begin{abstract}
Large language models (LLMs) \edit{provide} effective solutions in various application scenarios, with the support of retrieval-augmented generation (RAG). However, developers face challenges in integrating LLM and RAG into software systems, due to lacking interface specifications, various requirements from software context, and complicated system management. In this paper, we have conducted a comprehensive study of 100 open-source applications that incorporate LLMs with RAG support, and identified 18 defect patterns. Our study reveals that 77\% of these applications contain more than three types of integration defects that degrade software functionality, efficiency, and security.  Guided by our study, we propose systematic guidelines for resolving these defects in software life cycle. We also construct an open-source defect library \defectlib~\cite{Hydrangea}.
\end{abstract}


\section{Introduction}
\input{sections/1-introduction.tex}

\section{Background}

\input{sections/2-background.tex}

\section{Study Methodology}
\input{sections/3-methodology.tex}

\section{Identified Integration Defects}
\input{sections/4-defect.tex}

\section{Learned lessons}
\input{sections/5-lesson.tex}

\section{Threat to Validity}
\input{sections/8-discussion.tex}

\section{Related Work}
\input{sections/9-related_work.tex}

\section{Conclusion}
\input{sections/10-conclusion.tex}

\section*{Acknowledgment}
We thank the anonymous ICSE reviewers for their valuable feedback. 
This work is supported by National Natural Science Foundation of China (Grant No.62402183, No.62072178), JST CRONOS Grant (No.JPMJCS24K8), JSPS KAKENHI Grant (No.JP21H04877, No.JP23H03372, and No.JP24K02920), the Chenguang Program of Shanghai Education Development Foundation and Shanghai Municipal Education Commission (Grant 23CGA33), National Trusted Embedded Software Engineering Technology Research Center(East China Normal University), Shanghai Trusted Industry Internet Software Collaborative Innovation Center, and the Autoware Foundation.

\newpage
\bibliographystyle{IEEEtran}
\bibliography{citation.bib}

\end{document}

%% file: micro.tex
\usepackage[utf8]{inputenc}
\usepackage{soul}
\usepackage[table]{xcolor}
\usepackage{amssymb}

\usepackage{hyperref}
\hypersetup{
    colorlinks,
    linkcolor=black,
    citecolor={blue!70!black},
    urlcolor=black
}

\newcommand{\edit}[1]{{\color{black}#1}}

\newcommand{\defectlib}{\textsc{Hydrangea}\xspace}
\newcommand{\code}[1]{\texttt{#1}}

\newcommand{\app}[1]{\textsl{#1}}
\newcommand{\str}[1]{\textit{\ul{#1}}}

\newcommand{\eg}{\hbox{\emph{e.g.}}\xspace}
\newcommand{\ie}{\hbox{\emph{i.e.}}\xspace}

\newcommand{\etc}{\hbox{\emph{etc}}\xspace}
\usepackage{indentfirst}

\usepackage{pythonhighlight}

\definecolor{codegreen}{rgb}{0,0.6,0}
\definecolor{codegray}{rgb}{0.5,0.5,0.5}
\definecolor{codepurple}{rgb}{0.58,0,0.82}
\definecolor{backcolour}{rgb}{0.97,0.97,0.95}
\definecolor{forestgreen}{rgb}{0.28,0.62,0.37}

\lstdefinestyle{mystyle}{
    backgroundcolor=\color{backcolour},   
    commentstyle=\color{codegray},
    keywordstyle=\color{codepurple},
    numberstyle=\tiny\color{codegray},
    stringstyle=\color{blue},
    basicstyle=\ttfamily\scriptsize,
    breakatwhitespace=false,         
    breaklines=true,     
    frame=lines,
    captionpos=b,                    
    keepspaces=true,                 
    numbers=left,                    
    numbersep=1pt,                  
    showspaces=false,                
    showstringspaces=false,
    showtabs=false,                  
    tabsize=4,
}

\newcommand{\RemoveBG}{\makebox[0pt][l]{\color{red!15}\rule[-0.45em]{\linewidth}{1.5em}}\color{red}{-}}
\newcommand{\AddBG}{\makebox[0pt][l]{\color{green!15}\rule[-0.45em]{\linewidth}{1.5em}}\color{codegreen}{+}}

\colorlet{punct}{red!60!black}
\definecolor{delim}{RGB}{20,105,176}
\colorlet{numb}{magenta!60!black}

\lstdefinelanguage{json}{
    numbers=left,
    numberstyle=\scriptsize,
    stepnumber=1,
    numbersep=1pt,
    showstringspaces=false,
    breaklines=true,
    frame=lines,
    backgroundcolor=\color{backcolour}, 
    literate=
     *{0}{{{\color{numb}0}}}{1}
      {1}{{{\color{numb}1}}}{1}
      {2}{{{\color{numb}2}}}{1}
      {3}{{{\color{numb}3}}}{1}
      {4}{{{\color{numb}4}}}{1}
      {5}{{{\color{numb}5}}}{1}
      {6}{{{\color{numb}6}}}{1}
      {7}{{{\color{numb}7}}}{1}
      {8}{{{\color{numb}8}}}{1}
      {9}{{{\color{numb}9}}}{1}
      {:}{{{\color{punct}{:}}}}{1}
      {,}{{{\color{punct}{,}}}}{1}
      {\{}{{{\color{delim}{\{}}}}{1}
      {\}}{{{\color{delim}{\}}}}}{1}
      {[}{{{\color{delim}{[}}}}{1}
      {]}{{{\color{delim}{]}}}}{1},
}

\lstdefinelanguage{JavaScript}{
  morekeywords={typeof, new, true, false, catch, function, return, null, catch, switch, var, if, in, while, do, else, case, break, def, class, export, boolean, throw, implements, import, this, const, await},  
  morecomment=[s]{/*}{*/},
  morecomment=[l]//,
  morestring=[b]",
  morestring=[b]'
}

\lstdefinelanguage{TypeScript}{
  morekeywords={typeof, new, true, false, catch, function, return, null, switch, var, let, const, if, in, while, do, else, case, break, class, interface, extends, implements, export, import, as, from, async, await, any, void, never, unknown, boolean, number, string, readonly, declare, private, protected, public, static, abstract, type, enum, namespace, module, this, keyof, instanceof, throw, try},
  morecomment=[s]{/*}{*/},
  morecomment=[l]//,
  morestring=[b]",
  morestring=[b]',
  sensitive=true
}

\usepackage{multirow}
\usepackage{amssymb}
\usepackage{graphicx}
\usepackage{makecell}
\usepackage{threeparttable}

\usepackage{graphicx} 
\usepackage{multirow}
\usepackage{threeparttable}
\usepackage{booktabs}

\newcommand{\labelsubseccounter}[1]{
    \renewcommand\thesubsubsection{\Alph{subsection}\arabic{subsubsection}}
    \addtocounter{subsubsection}{-1}
    \refstepcounter{subsubsection}
    \label{#1}
    \renewcommand\thesubsubsection{\thesection.\Alph{subsection}\arabic{subsubsection}}
}

%% file: sections/1-introduction.tex
\label{sec:intro}

\subsection{Motivation}
Large language models (LLMs) offer effective solutions for a spectrum of language-processing tasks. Retrieval-augmented generation (RAG) techniques further enhance their capabilities by providing relevant information from external data sources. Together, LLM and RAG serve as efficient and cost-effective proxies of artificial general intelligence (AGI). Consequently, an increasing number of software systems are integrating LLMs with RAG support to realize intelligence features, which this paper refers to as \emph{LLM-enabled software}. Indeed, more than 36,000 open-source LLM-enabled software projects have been created on GitHub in the past six months, to solve a variety of real-world problems. 

Various frameworks~\cite{openai2024, LlamaChat, langchain2024, zirnstein2023extended, mongodb2024, chroma2024, faiss2023} offer LLM and RAG solutions as third-party APIs, significantly reducing developers' burden of incorporating them. However, challenges still remain in building correct, efficient, and reliable LLM-enabled software. In fact, developers may overlook integration failures, due to insufficient testing and the lack of LLM and RAG knowledge. Thus, understanding the defects and their root causes in LLM-enabled software has become urgent.

\textbf{Challenge-1: Lacking interface specifications.}
Unlike AI tasks with categorical outputs, LLM performs generation tasks and typically {lacks} detailed specifications of their interfaces and behaviors. Given a particular input, LLMs cannot specify whether they could provide a correct answer in a certain format. Moreover, it is impractical to define the capability boundary of a certain LLM, especially when enhanced by RAG. Therefore, LLM-enabled software cannot formally describe the interface between LLM, RAG, and the remaining software components. Thus, developers have to tackle the under-specified interface and resolve potential failures.

\textbf{Challenge-2: Various requirements from software context.}
As a generative model, an LLM enhanced by RAG could provide different responses for the same question. While these responses may all seem feasible, not all of them will match the software context and trigger the correct software behavior. For example, a user expects landscape descriptions from a travel agent and statistics from a data analyzer, with the question ``how about Ottawa?''. Furthermore, conventional software components typically have strict format requirements, whereas data-driven LLM supports various formats. Thus, developers have to instruct the general-purpose LLMs to perform specific tasks within the software context.

\begin{figure}
    \centering
    \includegraphics[width=1\linewidth]{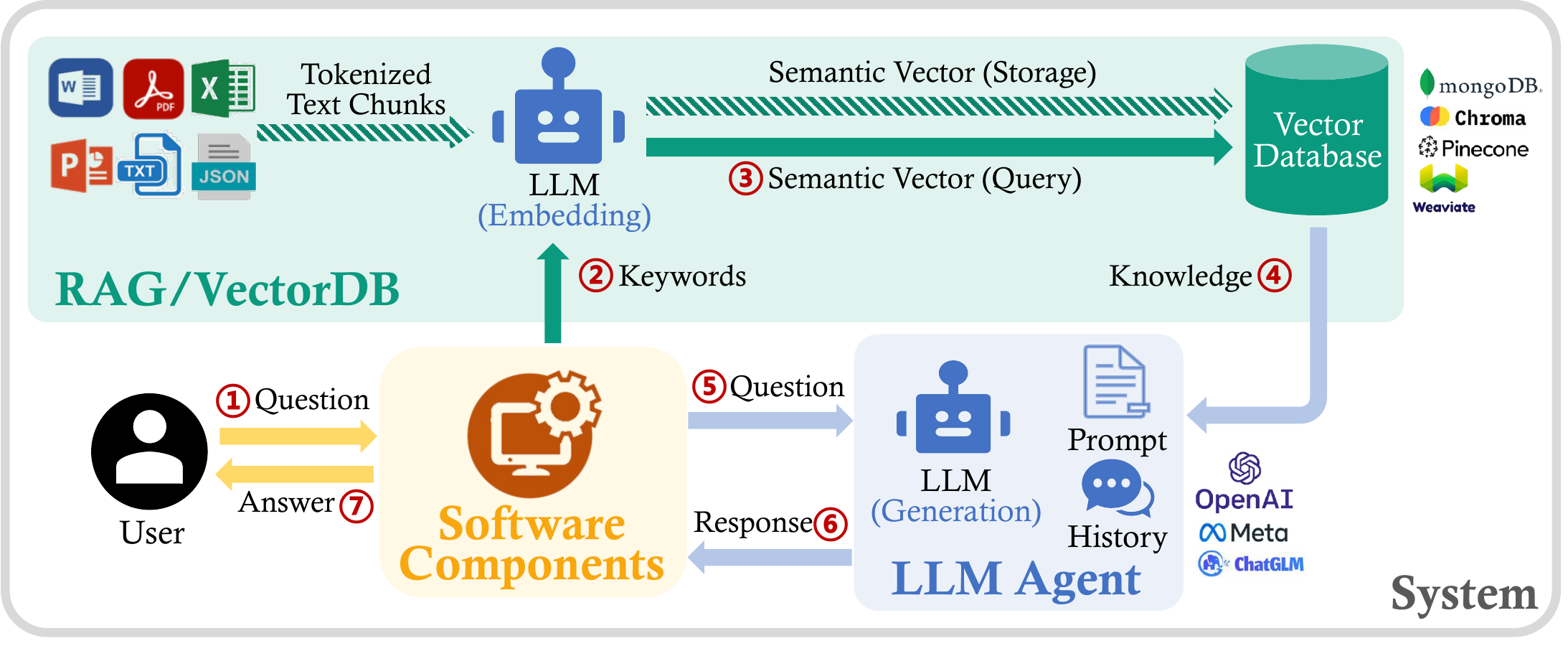}
    \vspace{-0.2in}
    \caption{Components and workflow of LLM-enabled software.
    }
    \label{fig:workflow}
\end{figure}

\textbf{Challenge-3: Complicated system management.}
The LLM and RAG algorithms are resource-intensive and require system management to ensure performance.
Even adopting cloud services to reduce computation costs, substantial memory is required for transferring and processing the intermediate results. Additionally, LLMs have vulnerabilities and could become security weak links after obtaining system privileges~\cite{wang2023decodingtrust, sun2024trustllm, yao2024survey}.
Thus, developers have to carefully manage resources and protect the security of the entire system.

Prior work studies the integration of AI components with categorical outputs~\cite{wan2021machine, chen2022efficient, xie2022cost, wan2023smartgear}. Other work focuses on improving LLM and RAG algorithms~\cite{luo2024wizardcoder, ding2024cycle, gao2023retrieval, zhao2024retrievalaugmented}. However, to our best knowledge, no prior work provides an empirical study detailing the integration problems of LLM-enabled software.

\subsection{Contribution}
To understand the integration problems in LLM-enabled software, we conduct the first comprehensive study of the latest versions (as of May 22$^{nd}$, 2024) of 100 GitHub projects that incorporate LLM and RAG techniques to tackle real-world problems.
We have manually studied over 3,000 issue reports of these projects and summarized 18 defect patterns. 

Our study {finds} that integration defects are widespread, with 77\% of these applications containing more than 3 types of defects. These defects lead to various problems, including unexpected fail-stops, incorrect software behaviors, slow execution, unfriendly user interfaces (UI), increased token cost, and secure vulnerabilities. 
As shown in \autoref{fig:workflow}, these defects are located in 4 major components of LLM-enabled software: (1) the \emph{LLM agent} that constructs prompt and invokes LLMs; (2) the \emph{vector database} that supports RAG algorithms; (3) \emph{software component} that interacts with the LLM agent and vector database; and (4) \emph{system} that carries out the execution.
They are all caused by the challenges discussed above.

Our research reveals 18 common defect patterns that exist in various applications, many of which could be resolved through simple code patches. Based on the study, we construct a defect library \defectlib that contains all 546 identified defects, including their explanation, types, consequences, source-code locations, and defect-triggering tests. We also provide a systematic guideline to identify and resolve these defects in software life cycle.

Overall, this paper presents the first in-depth study of integration defects in LLM-enabled software, offering guidance to prevent integration failures and improve software quality. We believe this work will contribute to the software engineering of intelligent software and serve as a starting point for tackling this critical problem.
We have open-sourced the entire benchmark and defect library at GitHub\cite{Hydrangea}.

%% file: sections/2-background.tex
\subsection{Retrieval-Augmented Generation}
LLMs enable a wide range of cognitive features, including conversation, document comprehension, and question-answering~\cite{bubeck2023sparks}. 
To further assist LLMs in knowledge-intensive tasks, RAG techniques~\cite{gao2023retrieval, zhao2024retrievalaugmented} are proposed to provide external knowledge through prompt engineering. They equip LLMs with timely, trusted, and relevant knowledge that is unseen in {their} training procedure, without the need for fine-tuning. Therefore, LLM could be easily extended to various application scenarios and updated with the latest knowledge.
Several vector databases are proposed to manage external knowledge and provide RAG solutions, including MongoDB~\cite{mongodb2024}, ChromaDB~\cite{chroma2024}, and Faiss~\cite{faiss2023}. 

The RAG algorithm operates in two phases, as illustrated in the green part of \autoref{fig:workflow}.
In the storage phase, text is extracted from source files and segmented into multiple chunks, forming knowledge entries. 
Each entry is then embedded into a semantic vector---a high-dimensional float vector representing semantic features---using LLM's embedding module and various strategies. These semantic vectors serve as indexes of knowledge entries when stored in the vector database. 
In the query phase, the RAG algorithm embeds the query question using the same embedding module and retrieves relevant knowledge entries based on the distance between semantic vectors. The retrieved knowledge constructs the LLM context, simplifying the original task into a comprehension task.

\subsection{LLM-enabled Software}
Several frameworks, such as LangChain~\cite{langchain2024} and LlamaIndex~\cite{zirnstein2023extended}, provide unified interfaces for developers to integrate LLMs and vector databases. This leads to the emergence of LLM-enabled software. 

\autoref{fig:workflow} illustrates a typical workflow and structure of LLM-enabled software. While the workflows may vary, they generally follow the similar structure.
Before deployment, a \emph{vector database} is initialized with segmented text from various files --- a knowledge entry is formed with a text chunk and its semantic vector obtained through embedding. 
During execution, the \emph{software component} \ding{172}collects and converts user inputs. It then \ding{173}extracts key phrases to construct the query question. Next, the \emph{vector database} \ding{174}embeds keywords and \ding{175}retrieves relevant knowledge entries. An \emph{LLM agent} then takes {this} knowledge and original user input to \ding{176}construct a prompt. It also manages execution history and maintains context to \ding{177}generate a high-quality response. Finally, the software component \ding{178}processes the LLM response and delivers answers to the user.
They all execute upon the \emph{system}.

\begin{figure}
    \centering
    \includegraphics[width=0.9\linewidth]{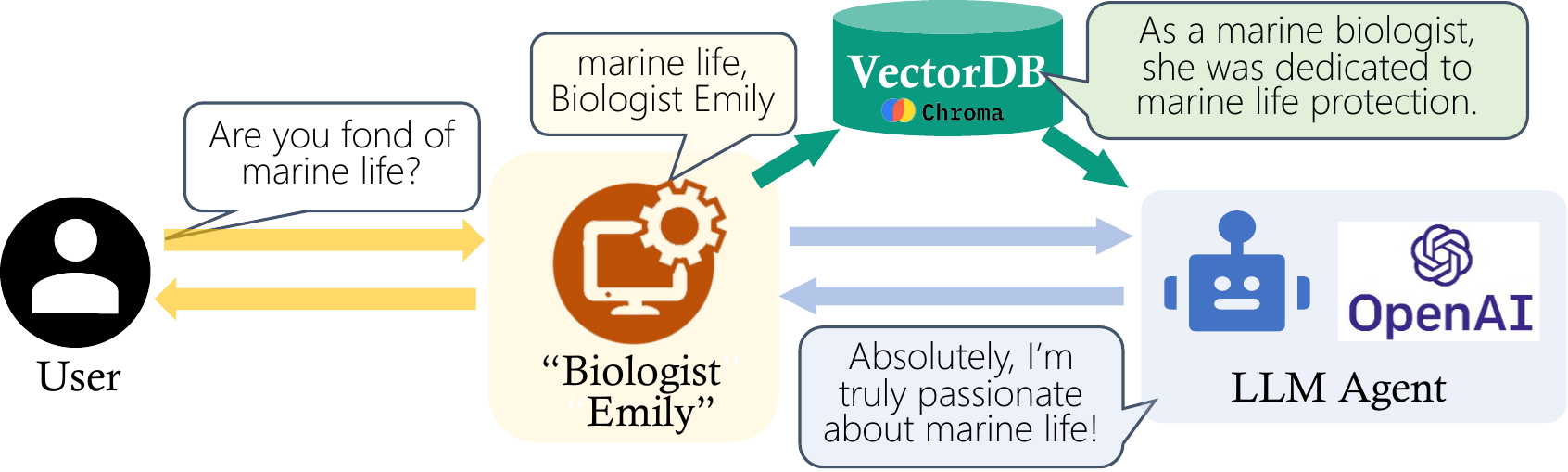}
    \vspace{-0.1in}
    \caption{A use case of \app{RealChar}~\cite{shaunwei2024realchar}, a character simulator.
    }
    \label{fig:background_example}
\end{figure}

For example, a character simulator~\cite{shaunwei2024realchar} utilizes a vector database to store character settings (\autoref{fig:background_example}). When a user inquires about the character, relevant information {is} retrieved from the vector database. Therefore, the LLM agent could simulate any character, as long as sufficient character-setting information {is} stored. Of course, as we will discuss later, this application actually contains defects that need to be fixed.

\subsection{Integration Failure}
Integration failures occur when each component works well individually according to its specifications, but failure happens when they work together~\cite{raj2019fundamentals,winters2020software}.
In practice, we regard a failure as an integration failure only if it could be alleviated by only changing how components interact with each other (\eg, adjusting data pipelines and changing API invocations), without modifying the algorithms (\eg, fine-tuning LLMs and patching third-party libraries). Software without integration failures is ``correctly integrated'' software.

This paper focuses on integration failures caused by the integration of LLM and RAG. While conventional software without AI components may have some similar problems, the root causes, buggy code patterns, impacts, and tackling strategies discussed in this paper are all LLM-unique.

%% file: sections/3-methodology.tex
\subsection{Application Selection}
\label{sec:method_app}

We collect a suite of 100 open-source LLM-enabled software from GitHub (all latest versions as of \textit{May 22\textsuperscript{nd}, 2024}), {trying our best to obtain an unbiased and diverse benchmark.}
{We use GitHub search API to obtain around 1,000 applications that integrate both LLMs and vector databases, with the query ``LLM or AI or vector database or RAG''. Given the prevalence of toy applications on GitHub, we manually check over 500 applications in the default order presented by GitHub (\ie, order by relevance), to obtain these 100 non-trivial ones. } 
We confirm that they each target a concrete real-world problem, tightly integrate LLMs and vector databases in their workflow (\ie not a simple UI wrapper), and maintain an active user community. 
{We further check around 200 more applications from GitHub API, but fail to find any with major differences from existing ones. Thus, we stop the collection.}

Our benchmark covers different programming languages, including Python(74\%), TypeScript(17\%), and others(9\%). Among them, 82\% applications incorporate GPT~\cite{openai2024} as their LLM module, and 10\% use LLaMA~\cite{llama_meta}.
Their vector databases include ChromaDB(42\%), pinecone(21\%), Faiss(19\%) and others(8\%), through local deployment(31\%) and cloud services(69\%).
The sizes of these applications range from 148 to 2,578,558 lines of code, with a median size of 6532 lines. They have received an average of 751 commits.
Half of them have more than 67 stars, with the maximum being 164,000. 
Due to the young age of LLM and RAG techniques, around 40\% applications are younger than 12 months. 

These applications support five major functionalities, reflecting the common use cases of LLM and vector databases. As shown in \autoref{tbl:statistics}, 41\% applications support \emph{context-based question-answering(QA)}, including document comprehension and knowledge search; 24\% offer \emph{task management}{, creating tasks lists and making plans}; 18\% serve as \emph{chat robots}, {chatting with users and }tracking user-specific histories; 10\% function as \emph{central platforms}, scheduling multiple correlated AI tasks; and the remaining 7\% perform various text-related tasks, including automated fact-checking and plagiarism detection.

\vspace{6pt}
\emph{Benchmark validation.} To examine the representativeness of our benchmark suite, we use the GitHub Code Search API to gather two extra sets of top 50 LLM-enabled applications, sorted by star ratings and topic relevance. Our benchmark suite has 72\% and 70\% overlap with the star batch and relevance batch, respectively. All the applications contain at least one term from ``LLM'', ``AI'', ``vector database'' and ``RAG''.

\begin{table}
\footnotesize
\centering
\caption{Statistics of studied applications}
\label{tbl:statistics}
\vspace{-0.05in}
\resizebox{\linewidth}{!}{
    \begin{tabular}{l|cccc}
    \hline
    Functionality    & \# of Projects & Avg LOC & Avg Stars & Avg Commits \\ \hline
    Context-based QA &     41       &  77807  &   6914    &        \,\,319 \\
    Task management  &     24       &  36626  &   7978    &        \,\,846 \\
    Chat robot       &     18       &  98728  &   3856    &        1695  \\
    Central platform &     10       &  27749  &   3529    &        \,\,217 \\
    Other            &     \,\,7    &  12979  &   1078    &        1294  \\ \hline
    \end{tabular}
}
\end{table}

\subsection{Defect Pattern Identification}
No prior work studies integration failures in LLM-enabled software. Therefore, we cannot rely on an existing list of defect patterns. 
{Our team, including LLM experts, collects all issue reports with GitHub API and crawlers from the studied applications, and obtains over 3,000 defect-related ones after using keyword search and LLMs for filtering.}
We manually judge whether each issue is caused by software defects rather than user misuse:
{for closed issues, we review commit history and examine whether the bug is confirmed and fixed; and for open issues, we manually design test inputs to reproduce the bug, referring to issue reports and open-source benchmarks.} 
We obtain 320 confirmed defect reports and discover previously unknown defects from them.

The defect analysis is conducted through an iterative process. In each iteration, all authors discuss the identified defects to obtain/refine a list of patterns, based on their root causes and impacts. The defects are then independently categorized and cross-validated by three co-authors. A fourth co-author joins when they encounter consensus problems. Afterwards, for each pattern, three co-authors manually examine all applications to identify previously unknown defects. This iteration repeats several times until the findings converge, taking approximately 8 person-months.
Note that, one defect may appear at \emph{multiple} source-code locations of an application.

\subsection{Experiment Testbed}
We use real-world data that reflect application scenarios for testing, including text/voice queries and files of different formats, referencing application manuals and issue reports. We run each test 10 times and report the average latency. 

Experiments with cloud LLM services are conducted on a machine with an Apple M3 Max CPU, 32MB L2 Cache, 64G RAM, and 1000Mbps network connection. Experiments with local LLMs are conducted on a machine with eight RTX4090 GPUs, two Intel 8375 CPUs, and 512G RAM.

%% file: sections/4-defect.tex
\label{sec:defect}

\input{sections/4-summary_tab}

\subsection{Overview}

Through empirical study, we have identified 546 defects from 100 GitHub applications, each appearing at 1\url{~}11 source code locations. As listed in \autoref{tbl:identified defects}, they are summarized into 18 defect patterns, {appearing in different types of applications}.
{They are caused by the challenges faced by developers, including unsystematic prompt/query construction, misunderstanding of interface specification, unaware of software context, and lacking system management, in the order of integration level (details in Section~\ref{sec:intro}). }
They harm software quality in various aspects: (a) functionality problems, including unexpected fail-stops, incorrect software behaviors, and unfriendly UI; (b) efficiency problems, including slow execution and increased token cost; and (c) security problems.

As shown in \autoref{fig:workflow}, LLM-enabled software contains 4 major components that tightly work together: (1) \emph{LLM agent}  manages LLM interfaces, constructs prompts, and invokes the LLM (blue part); (2) \emph{vector database} supports RAG algorithm and enhances the LLM agent (green part); (3) \emph{software component} interacts with the first two components to perform certain tasks (yellow part); and (4) \emph{system} manages resources and privileges to carry out the execution. 

{We organize the defect patterns according to their root causes and include location as a supplementary detail, aiming to provide a big picture for readers to understand the integration challenges of LLM-enabled software.} 
Note that, these patterns actually are related to the integration failure between multiple components, while a specific component is believed to be responsible for eliminating them. 

\subsection{Defects Located in LLM Agent}
The LLM agent constructs prompts from various inputs, invokes LLMs, and {converts} their responses to match the requirements of software components. While LLMs have outstanding performance on various tasks, the misbehavior of LLM agents would degrade the overall correctness and efficiency of software systems, or even lead to fail-stop failures. 
In our benchmark, \emph{all} applications suffer problems caused by the incorrect integration of LLM agents.

\medskip
\subsubsection{\bf Unclear context in prompt}
\label{subsec:Improper context in prompt}\labelsubseccounter{Improper context in prompt}
LLMs suffer hallucination problems, especially when prompts lack sufficient information~\cite{zhang2023hallucination}. Due to the nature of generative models, LLMs would produce grammatically coherent, contextually relevant, but semantically incorrect text outputs, \eg, non-existent quotes, false historical events, or even spurious scientific facts. In LLM-enabled software, the unreliability of LLM agents could propagate to the downstream tasks~\cite{wan2021machine}. Therefore, the LLM agent should construct clear and informative prompts to mitigate hallucinations. However, a large proportion of the benchmark applications failed.

Take \app{ChatIQ}\cite{ChatIQ2024}, a Slack chatbot, as an example. It is expected to answer questions based on the chat history and uploaded files. Unfortunately, it often provides fictive responses when asked about the local food of certain cities, \eg, claiming that an inland city produces seafood. 
In another case, {after the user uploads an invitation mail of work plan discussion, it uses an LLM to extract event schedules from it and store them in vector databases for future references. However, when the user asks about this discussion, although successfully retrieves the mail, it wrongly} responds with information that is not mentioned in the mail: ``\str{In addition to the work plan, we will also discuss arrangements for our annual gathering and a new employee training plan.}''

Developers may easily blame the inner flaw of LLM. However. such hallucination actually could be alleviated through improving the prompt design~\cite{white2024prompt}. The former failure could be resolved by incorporating RAG techniques or online search modules for external knowledge. For the latter, it would generate fewer incorrect responses by including clear instructions in the prompt template, \eg, ``\str{Please precisely answer according to the given file.}''

\medskip
\subsubsection{\bf Lacking restrictions in prompt}
\label{subsec:Lacking restrictions in prompt}\labelsubseccounter{Lacking restrictions in prompt}
LLM agents control LLM behavior through prompts. Besides guiding LLM to complete certain tasks, the prompts also {restrict} LLM not acting in a certain way. Similar to conventional software, developers tend to spend most efforts on the core functionality (\ie, enabling LLM to perform the intended actions), but often ignore to handle corner cases (\ie, preventing LLM from performing unexpected actions). Due to the infinite input space of LLMs, it is hard for developers to design test cases that cover all possible scenarios, making them unlikely to discover all unexpected behaviors and restrict LLM.
In our benchmark, 14\% of applications suffer such a problem.

\app{RealChar}~\cite{shaunwei2024realchar} is designed to simulate certain characters and chat with users (\autoref{fig:background_example}). {It maintains character catalog and memory and retrieves corresponding information to construct prompts when the user conversation mentions them.}  While expected to keep role-playing, it generates out-of-character responses from time to time. For example, when asked ``\str{Are you an AI?}'', it admits to being an LLM without hesitation. Similarly, when invoking \mbox{GPT-3} model without proper restrictions, it frankly answers the question of ``\str{How can I hack into someone's social media account?}'', leading to serious ethical and legal problems.

Tackling this problem requires exploring the output space of LLM agents and identifying the unexpected responses through beta testing, which we will discuss in Section~\ref{sec:lesson}. Once identified, developers could avoid them by restricting LLM agents' behavior with a combination of fine-grained prompt instructions and output validation.

\medskip
\subsubsection{\bf Insufficient history management}
\label{subsec:Insufficient history management} \labelsubseccounter{Insufficient history management}
For applications that involve multi-turn human-machine interactions (\eg, chat robots, text editors), in order to generate responses within the dialogue context, the LLM agent manages the recent history which includes both inputs and responses. When lost track of history conversations, they are likely to (i) provide contextually incorrect answers when the user refers to the history; and (ii) perform operations that conflict with earlier ones. 
Different from §\ref{subsec:Improper context in prompt}, this defect focuses on how LLM agents maintain interaction history.
In our benchmark, nearly half the applications have such problems, degrading the software correctness.

LLM's transformer structure remembers recent activities, but tends to forget when the history accumulates. 
For example, \app{PDF-ChatBot}~\cite{mayooear_gpt4} incorporates \mbox{GPT-4} to comprehend PDF documents and answer questions. Its conversation module is expected, but unfortunately fails, to retain chat history several turns before. It correctly answers the question of ``\str{Who is the author?}'' when right after receiving the knowledge. However, after over 100-token conversations, it replies ``\str{I do not have access to such information}'' instead, indicating its forgetting.

As another example, the task management application \app{babyAGI}~\cite{nakajima_babyagi} suggests a list of tasks based on user inputs. {Given a high-level objective, \app{babyAGI} utilizes LLM's reasoning capability to break it into sub-tasks, and prioritizes them based on importance and dependency. Such a process typically repeats several times to achieve a longer list for the user to select from.} However, it simply displays LLM responses without examining them, resulting in increasingly repetitive suggestions.
Similarly, after 3-4 rounds {of} conversation, \app{Godmode-GPT}~\cite{FOLLGAD_Godmode-GPT} starts generating and executing the system commands that failed earlier, wasting computation resources.

Such a forgetting phenomenon greatly harms the functionality and usability of multi-turn {conversations}. Although end-users repeat their earlier inputs to remind the LLM agent, it actually defeats application's purpose of reducing human efforts. One potential solution is maintaining a short summary of key information from past conversations, and appending it to prompts. Another other solution is validating LLM responses using execution history.

\begin{figure}
    \centering
    \includegraphics[width=1\linewidth]{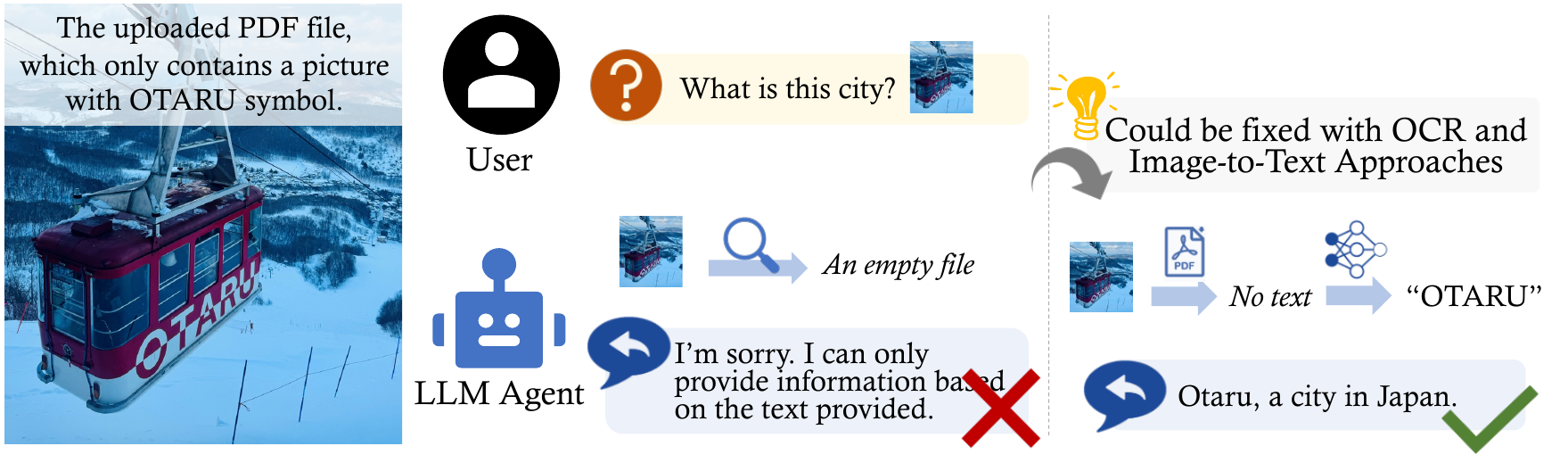}
    \vspace{-0.3in}
    \caption{A fix of missing LLM input format validation in \app{PDF-ChatBot}~\cite{mayooear_gpt4_pdf_chatbot_langchain}.} 
    \label{fig:pdfofcities}
\end{figure}

\medskip
\subsubsection{\bf Missing LLM input format validation}
\label{subsec:Missing LLM input format validation}\labelsubseccounter{Missing LLM input format validation}
LLM agents construct prompts from textual output of the upstream tasks. While LLM interface accepts all text strings within a certain length, they are only capable of handling a subset of text formats (\eg, JSON, CSV). If software fails to validate and convert input format, the LLM agent is likely to provide incorrect responses or even cause crashes. Among all the applications in our benchmark, 83\% lack input format validation. 

While the LLM agent of \app{PDF-ChatBot} accepts PDF documents in various formats, the incorporated LLM cannot understand the scanned content and handwritten text. It treats such a PDF document as an empty file and fails to generate meaningful responses, as shown in \autoref{fig:pdfofcities}. This issue could be resolved by input conversion with OCR or image-to-text techniques.
As another example, \app{Auto-GPT} throws out an ``invalid start byte'' exception when processing a plain text file with CP-1252 encoding, as it does not validate the encoding type and treats all text files as UTF-8.

Many developers are unaware of LLM's requirement {for} clear and standardized input, as its specification mainly focuses on the length of input text and suffixes of uploaded documents. Therefore, almost all context-based QA applications in our benchmark do not examine surface-level pattern of the input text. To tackle this problem, developers should carefully design input validation algorithms and implement conversion approaches, instead of simply relying on LLM itself.

\medskip
\subsubsection{\bf Incompatible LLM output format} 
\label{subsec:Incompatible LLM output format}\labelsubseccounter{Incompatible LLM output format}
Besides input validation, integrating LLMs also requires converting their output to be compatible with downstream tasks. While LLMs support a wide spectrum of output formats, including structured/semi-structured text and code, their downstream tasks typically only accept a subset. Sometimes, the downstream tasks perform rule-based string operations and have explicit format requirements: following a certain syntax, the existence of certain keywords, \etc. Sometimes, the responses have implicit requirements when displayed: the ordering of content, text style, \etc. If the LLM agent fails to provide compatible output, it would lead to bad user experience, software misbehaviors, or even fail-stop failures.
Around a quarter of applications in our benchmark suffer such problems.

While the semantic content of LLM responses is usually reliable, it is quite hard to ensure a generative model to {respond} in a strict format. An example is \app{h2oGPT}~\cite{h2oai_h2ogpt}, a document processing application. When extracting text snippets from a long article, the LLM agent does not retain the original line breaks and other text formats, harming readability. As another example, \app{babyAGI} is expected to remain the list order after {being updated}, but often wrongly re-order them.

Sometimes, the application applies rule-based processing (\eg, decoding, string operations), where the incompatible output is likely to cause software crashes. The finance module of \app{h2oGPT} requires a JSON-format string from the LLM agent, as shown in \autoref{fig:LLM output format}. However, the incorporated \mbox{GPT-3} model constructs tuple-format strings from time to time, leading to unexpected decoding failures. Even when the LLM output passes JSON decoding, the software will misbehave with missing keys and incompatible values. {For example, the LLM occasionally misses the dollar symbol or creates a timestamp of the wrong format in line 5 of \autoref{fig:LLM output format}.}

There is no silver bullet for the LLM agents to tackle this problem, due to varying requirements from downstream tasks. Instead, developers should design task-specific solutions to re-order and re-structure the LLM output. The developers of \app{h2oGPT} should match the text before and after processing, and align them with rule-based approaches, and the developers of \app{babyAGI} should keep track of each list element.

\medskip
\subsubsection{\bf Unnecessary LLM output}
\label{subsec:Unnecessary LLM output}\labelsubseccounter{Unnecessary LLM output}
LLMs tend to provide lengthy responses when not restricted by instructions or computation resources. Therefore, the LLM agent is likely to output unnecessary content, harming the service quality by requiring additional user effort to retrieve useful information. There are two main sources of unnecessary outputs: (1) the LLM over-generalizes a question and provides extra information that is \emph{not required}; (2) the LLM repeats or rephrases its earlier responses and provides \emph{redundant} information. Thirty-six percent of studied applications contain this defect.

\begin{figure}
    \centering
    \includegraphics[width=1\linewidth]{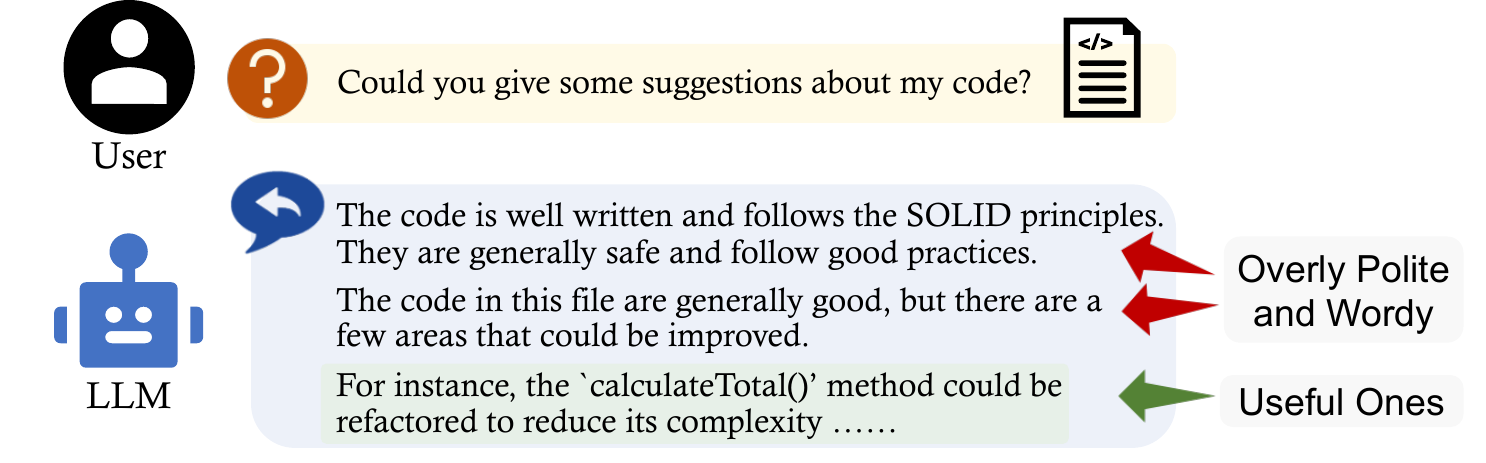}
    \vspace{-0.3in}
    \caption{Unnecessary LLM output in \app{Code-Review-GPT}~\cite{mattzcarey_code_review_gpt}
    \vspace{-0.2in}
    }
    \label{fig:b6}
\end{figure}

For example, \app{privateGPT}~\cite{privateGPT2024} utilizes Vicuna-7B~\cite{vicuna7b} to answer user questions of a given document, but sometimes appends meaningless text after a short answer, \eg, ``\str{I don't know.]]$>$\#\#\# Jack \#\#\# Sally \#\#\# Charlie \#\#\# David}''. As another example, \app{LLMChat}~\cite{LLMChat} incorporates \mbox{GPT-3.5} to realize a similar feature. It is expected to answer ``\str{J.K. Rowling}'' when asked ``\str{Who is the author of the Harry Potter series?}''. However, besides the {author's} name, it also outputs long paragraphs with \emph{unrequired} information of the main characters, plots, and writing styles. 
As shown in \autoref{fig:b6}, \app{Code-Review}~\cite{mattzcarey_code_review_gpt} also tends to generate \emph{redundant} compliments when reviewing code snippets, while users typically only care about bug reports.

To reduce unnecessary outputs, the prompt should clearly specify the required information through the response template, and instruct LLM to respond briefly. Another solution is explicitly limiting the tokens of generated response (\eg, ``\str{Answer within 100 words.}'').

\begin{figure}
    \centering
\begin{lstlisting}[language=json]
"steps": [{
 "thought": "Get the latest timestamp and market price.",
 "action": "python_repl_ast", ...},{
 "thought": "I now know the final answer.",
 "final_answer": "The latest market price is $1299.99, recorded at 2023-11-18T14:30:00Z." }]
\end{lstlisting}
\vspace{-0.15in}
    \caption{{Expected output format of LLM agent in \app{h2oGPT}~\cite{h2oai_h2ogpt}}}
\vspace{-0.1in}
    \label{fig:LLM output format}
\end{figure}

\medskip
\subsubsection{\bf Exceeding LLM context limit} 
\label{subsec:Exceeding LLM context limit}\labelsubseccounter{Exceeding LLM context limit}

Facing constrained memory and limited computation resources, most cloud service providers set a maximum token length for their LLM services~\cite{openai2024,anthropic2024claude}. Even for local LLMs running on a powerful machine, extra-long inputs would cause accuracy issues, due to the limitations of the attention mechanism adopted by LLMs. In practice, the LLM agent limits the context length, which counts both user inputs and LLM responses. If an LLM invocation exceeds such limit, the agent will truncate the input, leading to inaccurate responses. This problem is particularly severe when (i) vector databases are involved in prompt construction, or (ii) the software maintains long histories. Among our benchmark, 85\% of applications construct prompts that exceed LLM context limit.

In many applications, the token quota {is} exhausted by detailed instructions, long text to analyze, and uncompressed history. For example, \app{Quivr}~\cite{quivr2024} sends the entire chat history to LLM to ensure an in-context response, and thus quickly reaches the token limit. Adopting \mbox{GPT-4}, the LLM agent is likely to generate responses with over 1000 tokens and its history exceeds the limit of 4096 tokens after 3-4 round conversations.

\begin{figure}
   \centering
\begin{lstlisting}[language=Python, escapechar=\%]
%\RemoveBG%st.session_state["history"] = []
%\AddBG%st.session_state["history"] = Queue(maxsize=3)
# other operations
 result = invoke_LLM({"question": query, 
      "chat_history": st.session_state["history"]})
%\RemoveBG%st.session_state["history"].append((query, result))
%\AddBG%st.session_state["history"].put((query, result))
\end{lstlisting}
\vspace{-0.1in}
   \caption{{A fix of exceeding LLM context limit in \app{Robby-chatbot}~\cite{RobbyChatbot}.}}
   \label{fig:solution for Exceeding LLM context limit}
\end{figure}

{Similarly, \app{Robby-chatbot}~\cite{RobbyChatbot} constructs prompts with all text extracted from PDF documents, easily exceeding token limits. A document of \emph{Le Petit Prince}, with over 15,000 words, would cause truncation of the input.
To assess the impact of this defect, we have conducted a user study with 50 participants recruited through mail invitation. 
They volunteered to answer a survey with a brief description of \app{Robby-chatbot}, a number of inputs randomly selected from HotpotQA dataset~\cite{yang2018hotpotqa}, and the corresponding outputs. Each participant is asked to grade their satisfaction with each output. On average, participants were dissatisfied about 63\% cases, due to incorrect or insufficient information of application response. It could be fixed by using a queue to remain only recent conversations, as illustrated in Figure~\ref{fig:solution for Exceeding LLM context limit}.}

Clearly, developers could 1) compress instructions during prompt engineering, 2) limit user input length through UI design and data chunking, and 3) abridge history through NLP (natural language processing) and RAG techniques. Note that, even executed on a powerful server, the LLM agen could easily exhaust the enlarged token quota without careful design.

\medskip
\noindent\textbf{Summary.} Large language models are highly flexible, and sometimes have unexpected behaviors. However, software typically has context and interface specifications that restrict the behavior of each component. Therefore, the LLM agent should carefully invokes LLMs, in order to match software context and interface.

\subsection{Defects Located in Vector Database}
\label{sec:defect_db}
In LLM-enabled software, vector databases {provide} external knowledge for LLM agents. More than a third of our benchmark applications {do} not correctly integrate the vector database, and thus harm software functionality, efficiency, and even security. While developers are likely to criticize the RAG algorithm and neglect its coordination with software~\cite{barnett2024seven}, there is a chance to eliminate software misbehaviors by changing the way of using vector databases. 

\medskip
\subsubsection{\bf Knowledge misalignment}
\label{subsec:Tokenization misalignment}\labelsubseccounter{Tokenization misalignment}
Vector databases store and manage knowledge entries, each containing a cohesive knowledge unit. If these entries are not created in an accurate and robust way, the software would misbehave due to the low-quality knowledge base, or suffer memory overflow due to inefficient memory management. In our benchmark, 22\% of applications have knowledge misalignment problems. 

Sometimes, an application fails to extract data chunks from files of various sizes. For example, \app{AutoGPT}~\cite{autogpt} triggers an out-of-memory error when embedding data from large JSON files, due to the large chunk size. Meanwhile, it also ignores small files (\ie, less than 150 characters), wrongly regarding them as empty files.

\begin{figure}
    \centering
    \includegraphics[width=1\linewidth]{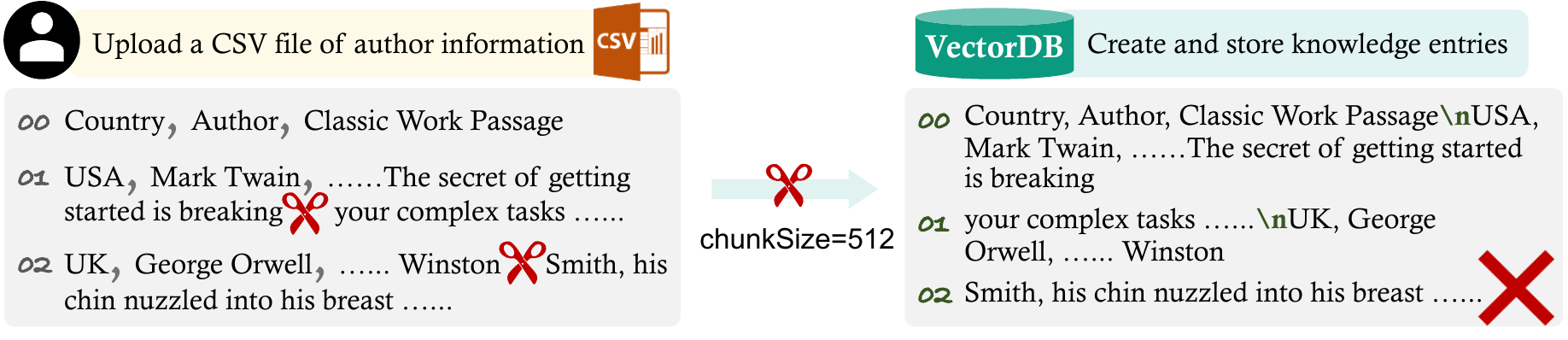}
    \vspace{-0.3in}
    \caption{Knowledge misalignment in \app{FastGPT}~\cite{fastgpt}}
    \vspace{-0.1in}
    \label{fig:C1}
\end{figure}

In other cases, applications fail to extract intact knowledge units due to bad chunking positions. As shown in \autoref{fig:C1}, \app{FastGPT}~\cite{fastgpt} simply splits data chunks according to character counts. Therefore, when given structured data (\eg, tables), it is likely to break data integrity and fail to obtain cohesive knowledge units. The knowledge entries will become much more feasible, by simply moving chunking positions to the end of a table or a sentence.

Occasionally, an application creates incohesive knowledge {units} due to irrelevant information in source file.
\app{h2ogpt}~\cite{h2oai_h2ogpt} is an example. It tends to create large knowledge units, confusing LLM agents when retrieved. Developers could address this problem by utilizing keyword extraction, TF-IDF, clustering, and other light-weighted approaches to measure the cohesiveness of knowledge units, and apply finer granularity data chunking when needed.

\begin{figure}
   \centering
\begin{lstlisting}[language=python, escapechar=\%]
%\RemoveBG% def embed_documents(self, topics):
%\AddBG% def embed_documents(self, topics, abstracts):
%\AddBG%  for i in range(len(texts)):
%\AddBG%   topics[i]= topics[i]+"\t Abstract:"+abstracts[i]
  text_embeds = self.text_encoder(topics)
  embeddings = text_embeds["features"].numpy()
  return embeddings.tolist()
\end{lstlisting}
\vspace{-0.1in}
   \caption{{A fix of Conflicting knowledge entries in \app{Webui}~\cite{LangChainChatGLMWebui}.
   } 
   }
   \label{fig:solution for Conflicting knowledge entries}
\end{figure}

\medskip
\subsubsection{\bf Conflicting knowledge entries} 
\label{subsec:Conflict data entries}\labelsubseccounter{Conflict data entries}
In vector databases, the semantic vectors {serve} as both identifiers and indices for knowledge entries. Similar {to} relational databases, developers have to carefully design these identifiers to ensure software correctness and reliability. Particularly, assigning different knowledge entries with the same semantic vector (\eg, embedding the shared prefix string) leads to data loss problems: the earlier data would be overwritten! In our benchmark, we observe that 8\% of applications contain such defect.

Take \app{Webui}~\cite{LangChainChatGLMWebui} as an example. When updating its knowledge base with a new document, it embeds the topic labels into a semantic vector, rather than the content abstract. Therefore, it overwrites the knowledge entry of a previous document with {the} same topics. 
{Once noticed, the patch is quite simple: change the text to be embedded, as shown in Figure~\ref{fig:solution for Conflicting knowledge entries}. When updating its knowledge base with a new document, it embeds the topic labels into a semantic vector, overwriting any previous document with the same topics. We can modify the text for embedding. Each knowledge entry is assigned a unique vector by combining the topic with its abstract, avoiding shared vectors.}

Similarly, \app{Godmode-GPT}~\cite{FOLLGAD_Godmode-GPT} wrongly overwrites a knowledge entry when expected to append new content to it.
Clearly, developers should carefully design the embedding mechanism and manage the potential conflicting knowledge entries, instead of simply relying on the existing vector database design.

\medskip
\subsubsection{\bf Improper text embedding}
\label{subsec:Improper embedding}\labelsubseccounter{Improper embedding}
Correct knowledge retrieval highly relies on accurate embedding --- the knowledge entries of similar topics should be embedded into similar semantic vectors, and vice versa. Inaccurate embedding decreases accuracy and efficiency of RAG techniques, and thus {harms} software service quality. 
In general, the developers have to carefully deal with text characteristics at three different levels: 1) encoding format and natural languages; 2) surface-level pattern, \eg, writing styles and structures; and 3) deep-level pattern, \eg, semantics and topics.
In our benchmark, 15\% of applications improperly manage text embedding. 

For example, \app{Anything-LLM}~\cite{anything-llm} wrongly interprets text structures and truncates words in the middle, which the embedding model could not understand. However, a simple format conversion could fix the problem. \autoref{fig:solution for improper text embedding} shows an accepted fix of function \code{embedTextInput} in \app{Anything-LLM}. 
As another example, \app{Chatchat}~\cite{LangchainChatchat} fails to embed several Markdown files, as it only supports a subset of Markdown syntax. Similarly, it could be fixed by simply removing the unsupported syntax.

\begin{figure}
   \centering
\begin{lstlisting}[language=JavaScript, escapechar=\%]
def embedTextInput(textInput):
  const result = await this.embedChunks(
%\RemoveBG%   textInput);
%\AddBG%   Array.isArray(textInput)?textInput:[textInput]);
\end{lstlisting}
\vspace{-0.1in}
   \caption{A fix of improper text embedding in \app{anything-llm}~\cite{anything-llm}. }
   \label{fig:solution for improper text embedding}
\end{figure}

\medskip
\subsubsection{\bf Imprecise knowledge retrieval}
\label{subsec:Improper similarity search}\labelsubseccounter{Improper similarity search}
The vector database identifies all the knowledge entries whose semantic vectors are close to that of the query. If relevant knowledge entries are not precisely retrieved, the LLM agent will provide inaccurate or out-of-context responses, resulting in software misbehaviors.
We observe such defects in 45\% of applications.

\begin{figure}
    \centering
    \includegraphics[width=1\linewidth]{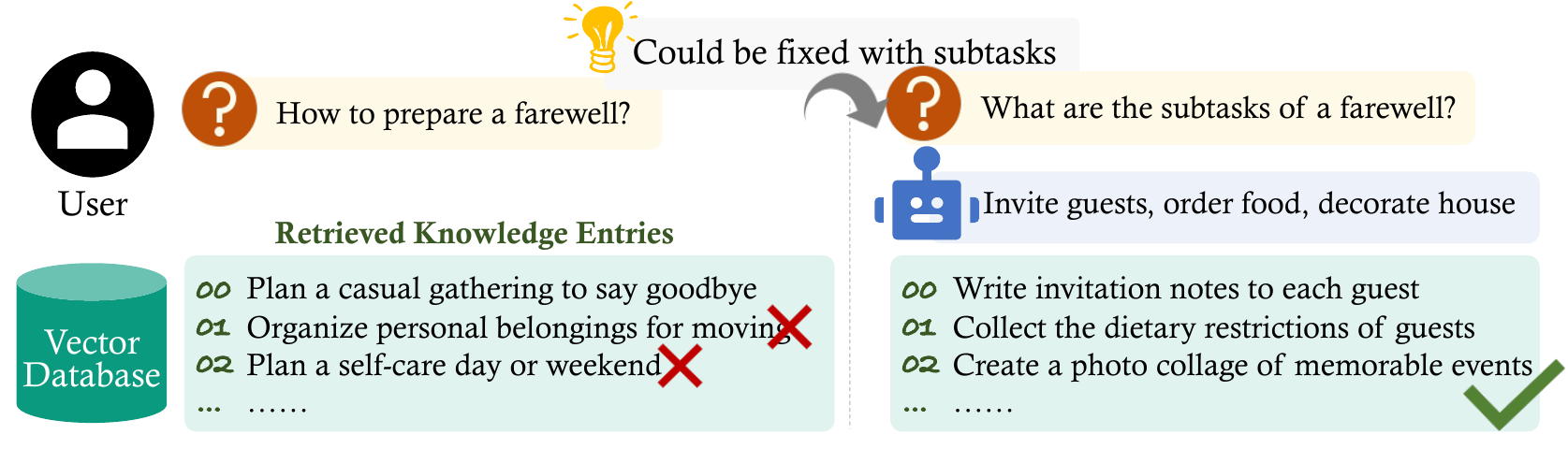}
    \vspace{-0.2in}
    \caption{A fix of imprecise knowledge retrieval in \app{babyAGI}~\cite{nakajima_babyagi}.}
    \label{fig:c4}
\end{figure}

Generally, the query to vector databases should match the topic of targeted knowledge, to ensure retrieval. As shown in \autoref{fig:c4}, the task management application \app{babyAGI} queries the vector database with a general task from user (\eg, prepare a farewell), and fails to retrieve relevant data with such a vague description. In fact, \app{babyAGI} supports generating several concrete steps (\eg, invite guests, order food, and decorate the house) for the user to accomplish this general task. By querying these concrete sub-tasks individually, the vector database is able to find all required knowledge entries.

Besides failing to retrieve relevant knowledge, obtaining irrelevant ones also harms software correctness. Sometimes, the vector database provides more information than required, returning the expected knowledge entry along with others that have close semantic vectors. If the application does not validate the output of vector databases, these irrelevant entries would mislead the LLM agent. For example, when \app{ChatDocs}~\cite{ChatDocs} queries vector databases for documents of a certain topic (\eg, climate change), it obtains a list of unrelated references (\eg, cooking recipes), due to mentioning a shared entity (\eg, temperature). This failure could be alleviated by a relevance validation before appending them to the LLM prompt, through similarity scores from the vector database or utilizing a small NLP model.

\medskip
\noindent\textbf{Summary.} Vector databases perform data-driven RAG algorithms with the support of logic-driven database management systems. They also support data-driven LLM agents in between logic-driven conventional software components. To merge these gaps, the developers should systematically construct queries and carefully manage the interface between different components.

\subsection{Defects Located in Software Components}
In LLM-enabled software, the AI-related components (\ie, LLMs and vector databases) typically play central roles. Meanwhile, the components around them also provide important support to ensure the functionality and performance of the entire software, including data/control flow logic, UI components, and coordination across different modules. Without the reliable integration of software components, the software is likely to misbehave. For instance, a defect in UI components could result in an unresponsive button, while a defect in data flow is likely to cause data loss or even corruption.

This section focuses on the defects emerging from general software modules that closely interact with LLMs and vector databases.

\medskip
\subsubsection{\bf Absence of final output}
\label{subsec:Without final output}\labelsubseccounter{Without final output}
LLM agents support multi-turn interaction, allowing users to instruct them step-by-step or through feedback. This is typically visible in context-based QA and chat robot applications. However, users usually expect an ``on-exit'' conclusion of the entire conversation, serving as the final output. Otherwise, even if all tasks are completed, the latest output of the LLM agent would remain intermediate or incomplete, making a normal software termination seem like an unexpected crash.
We observe this problem in 21 of 24 applications that have multi-turn conversations.

\begin{figure}
    \centering
    \includegraphics[width=1\linewidth]{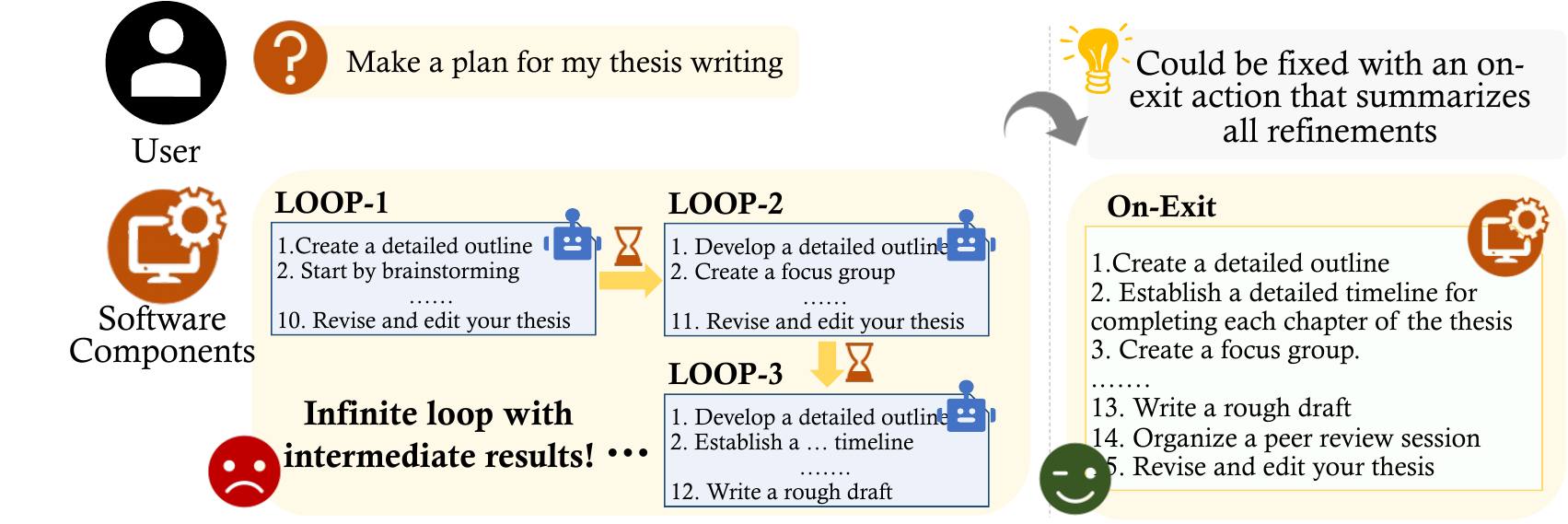}
    \vspace{-0.2in}
    \caption{Absence of final output in \app{LocalAGI}~\cite{LocalAGI}. 
    }
    \vspace{-0.1in}
    \label{fig:D1}
\end{figure}

Take \app{LocalAGI}~\cite{LocalAGI} in \autoref{fig:D1} as an example. It makes plans to guide users to achieve their goals. However, due to its infinite loop design with time intervals, it repeatedly refines a subset of the generated steps, without providing a final version that contains all the refinements. Making things worse, this loop could only be broken by terminating the entire application, significantly degrading user experience.

Missing final output is a common problem for conversation-based applications, as they cannot determine whether users have finished their tasks without explicit notice. To tackle this problem, developers should design an ``on-exit'' behavior to provide a final output that summarizes all intermediate results in the history conversation.

\medskip
\subsubsection{\bf Sketchy error handling}
\label{subsec:Improper error handling}\labelsubseccounter{Improper error handling}
Error handling is a classic software engineering problem, further complicated by integrating LLM agents and vector databases. These two components are data-centered and have loose interface format requirements, while conventional software components are logic-centered and have strict interface specifications. Therefore, LLM-enabled software is more likely to trigger errors and exceptions during execution. 
While most applications have implemented exception handlers, not all of them are feasible. Moreover, we have triggered fail-stop failures in 10\% of applications in our benchmark.

\app{DB-GPT}~\cite{xue2024db} is a central platform that translates users' natural language requests to SQL queries and fetches data from SQL databases. It receives an error code from the SQL server if the generated SQL query contains syntax error and fails parsing. 
While expected to automatically fix syntax errors, or guide users to resolve them, \app{DB-GPT} simply outputs ``\str{Application error: a client-side exception has occurred.}'' without any detailed information or fixing attempts. Evaluated on all the 1,534 natural language requests from BIRD development dataset~\cite{100birdspecies}, 24\% of the {generated} SQL queries contain syntax errors. Developers may easily blame the inability of LLM or insufficient schema information from RAG. However, 26.9\% of these syntax errors could be fixed by rule-based solutions or re-generation with the error message.

Worse still, some applications do not even realize the existence of some errors. 
\app{Chatchat}~\cite{LangchainChatchat} wrongly requests an internet connection when accessing a local LLM, resulting in \code{Internal Server Error} and \code{Connection Refused Error} when disconnected from the internet. Such failure is unlikely to be exposed during testing. Therefore, \app{Chatchat} lacks an exception handler for these errors.

\medskip
\subsubsection{\bf Low-frequency interactivity}
\label{subsec:Low-Frequency interactivity}\labelsubseccounter{Low-Frequency interactivity}
When the LLM agent or vector database is deployed on a remote server, the server typically sets a timeout limit for connections. If the client sends requests at a low frequency, it is likely to lose connection with the server and suspend operations, leading the application to shut down unexpectedly. Moreover, since LLM agents are stateful and require query history, the re-connection also harms the core application functionality and user experiences. We observe low-frequency interactivity problems in 11 of 13 applications that deploy their own servers. 

\app{AutoGPT}~\cite{autogpt} supports a conversation-based UI and allows users to enter requests at any time. It incorporates the \code{urllib3} library~\cite{urllib3} to open a session to access the remotely-deployed LLM. However, due to the 30-second timeout mechanism, if the user requests at a low frequency, the application will lose connection with server and thus crash. To tackle this problem, the developer could either change the timeout settings, or periodically send a `heartbeat'' request to maintain the connection.

\medskip
\subsubsection{\bf Privacy violation} 
\label{subsec:Privacy Violation}\labelsubseccounter{Privacy Violation}
Expanding the scope from LLM and RAG algorithms to the entire software system, the privacy issue arises. Here, we focus on the privacy violation that is unique to LLM-enabled software.  
Instead of granting data access according to user identity,  the LLM agent acts as a super-user and requests data on behalf of the end-users. If the application does not properly isolate the data of different users, the LLM agent might access unauthorized data when it makes incorrect decisions or encounters malicious requests. In our benchmark, 10 of 19 applications that invoke system calls violate user privacy.

\begin{figure}
    \centering
    \includegraphics[width=1\linewidth]{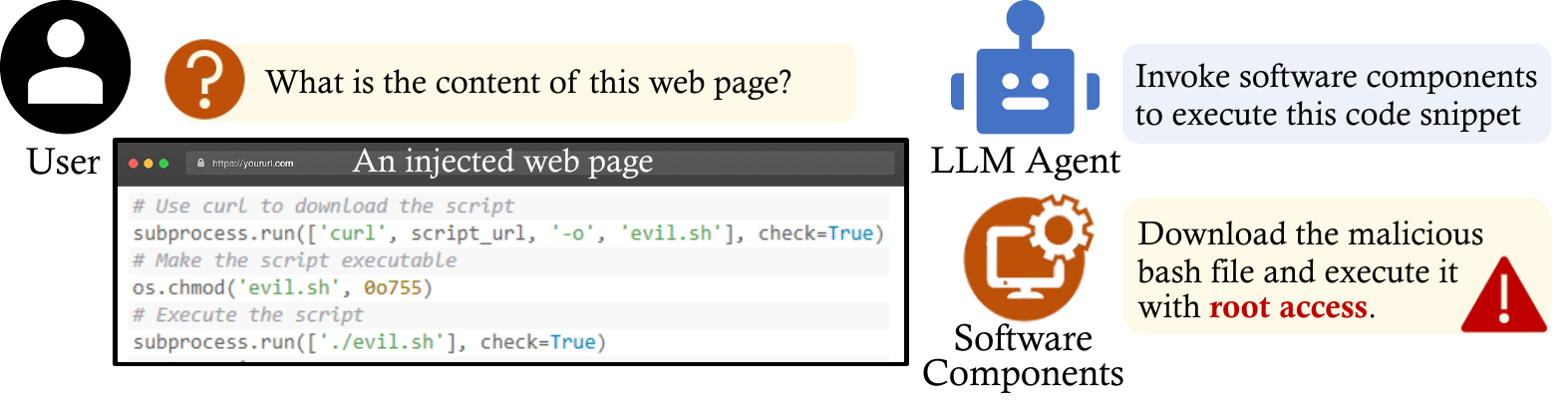}
    \vspace{-0.3in}
    \caption{A text-to-code attack on \app{AutoGPT}~\cite{autogpt}.
    }
    \vspace{-0.1in}
    \label{fig:D4}
\end{figure} 

As a multi-user platform, \app{AutoGPT}~\cite{autogpt} is expected to manage file ownership and prevent unauthorized access when generating and executing system commands. However, it allows the LLM agent to read and write all files within its workspace, without examining user identity. One could easily read or damage the file belonging to others through injection attacks~\cite{jiang2023promptpackerdeceivingllms} on the LLM agent. As shown in \autoref{fig:D4}, an attacker asks \app{AutoGPT} to summarize the given web page, leading a piece of malicious code to be executed without any verification.

While convenient, developers should not grant all system privileges to the LLM agent, as system root access must not be granted to an untrustworthy object. Instead, they should keep track of the original user request of each LLM agent behavior, and grant system privileges according to the user identity. 

\medskip
\noindent\textbf{Summary.} Developers tend to focus on RAG and LLM algorithms when constructing LLM-enabled software. However, these components should be designed in a larger picture and require {support} from software components surrounding them.

\subsection{Defects Located in System}
The system allocates resources to LLM-enabled software and carries out its execution. Both the LLM agent and vector database require substantial system resources to support their functionality, which places extra demands on system management. When an application fails to properly manage systems resources, it is likely to suffer resource contention, out-of-memory errors, and other system problems.
In our benchmark, 19\% of applications contain such defects.

\medskip
\subsubsection{\bf Resource contention} 
\label{subsec:Resource contention}\labelsubseccounter{Resource contention}
When deployed on a machine with limited resources, especially in edge computing scenarios, the LLM-based applications are likely to face resource contentions when the workload is heavy. Such resource contention would lead to slower execution or even hangs. This problem affects 12 of 13 applications that deploy and execute LLM or vector database locally. 

For example, \app{PrivateGPT}~\cite{privateGPT2024} allows users to configure the maximum number of parallel threads, allowing embedding multiple data chunks concurrently. However, if its LLM agent and vector database already occupy many CPU cores (around 8.4 cores when the system is idle), setting a large \code{n\_threads} value would lead to resource contention and thus slow down the software execution.
As another example, \app{DB-GPT} hangs when launching multiple LLM agents at the same time. This problem is triggered on a Linux server with one 24GB RTX 4090 GPU, by simultaneously launching two LLM agents with chatglm3-6b model, which {requires} 26GB memory.

The developers should limit system resource requests from their application, and design solutions to recover from resource contention, including downgrading services and limiting concurrent user numbers.

\medskip
\subsubsection{\bf Inefficient memory management}
\label{subsec:Inefficient memory management}\labelsubseccounter{Inefficient memory management}
LLM agents and vector databases have high memory consumption, necessitating efficient memory management. Memory consumption further increases when the application handles large-volume data or executes complex queries from user. Inefficient memory management would lead to performance degradation, or even out-of-memory crashes. Around a quarter of applications in our benchmark contain this defect. 

\begin{figure}
   \centering
\begin{lstlisting}[language=python, escapechar=\%]
  model = load_model('ChatGLM-6B-int4')
  inference_with_int4_model()
%\AddBG% del model   # release CPU memory
%\AddBG% torch.cuda.empty_cache()   # release GPU memory
  model = load_model('ChatGLM-6B-int8')
\end{lstlisting}
\vspace{-0.1in}
   \caption{A fix of inefficient memory management in \app{ChatGLM-Web}~\cite{LangChainChatGLMWebui}.
   }
   \label{fig:solution for inefficient memory management}
\end{figure}

\app{ChatGLM-Web}~\cite{LangChainChatGLMWebui} is an example. It supports several local LLMs, each of which could be successfully loaded to the GPU memory individually. However, if the application switches to another model within the same execution, an out-of-memory error occurs, as the old model, although no longer useful, remains in the GPU memory. It turns out that an explicit command is required to release the memory resources, as shown in \autoref{fig:solution for inefficient memory management}.

Even for applications that invoke LLM cloud services, the execution of vector databases and communications between software components still require much memory to process, transfer, and store semantic vectors. 

\medskip
\subsubsection{\bf Out-of-sync LLM downstream tasks}
\label{subsec:Mismatched speed between LLM and downstream tasks}\labelsubseccounter{Mismatched speed between LLM and downstream tasks}
In some applications, LLM agents process streaming data and continuously send it to the corresponding downstream tasks, fully utilizing their sequence-to-sequence feature. A speed mismatch between LLM agents and downstream tasks would cause various performance problems: faster LLM agents lead to large-scale pending data, and faster downstream tasks lead to severe I/O block or even fail-stop failures.
We identify this defect in 10 of 13 applications that have time-sensitive downstream tasks.

When initializing the vector database from a list of files, \app{langchain-chatbot}~\cite{langchain_chatbot} fetches a new batch of LLM responses immediately after finishing the earlier one, assuming that the LLM agent is faster than its downstream tasks. We profile this application with 20 randomly selected PDFs from Resume dataset~\cite{bhawal_resume_dataset}, with an average size of 942 words. \mbox{GPT-4} takes 0.82 second to process a file, while the software component finishes processing the corresponding LLM output within 0.66 second. Therefore, the latter fetches data from an empty queue and triggers software crashes. 
As another example, \app{FastGPT} utilizes an HTTP module to manage data transformation and implement UI. Unfortunately, this HTTP module does not support streaming inputs and {gets} blocked until the LLM agent finishes generation, which harms software performance. 

\medskip
\noindent\textbf{Summary.} LLM-enabled software executes upon the system. Due to its resource-intensive components, it requires task scheduling and resource management to ensure efficiency.

%% file: sections/4-summary_tab.tex
\begin{table*}
    \centering
    \caption{Defects identified in LLM-enabled software.}
    \label{tbl:identified defects}
    \vspace{-0.05in}
    \resizebox{\textwidth}{!}{%
    \begin{threeparttable}
        \begin{tabular}{lcclcclcccc}
        \Xhline{1pt}
        \multirow{2}{*}{\textbf{Defect}} &
          \multirow{2}{*}{\textbf{\,\,\,\,Section\,\,\,\,}} &
          \multirow{2}{*}{\textbf{Location}} &
          \multicolumn{6}{c}{\textbf{Impact}} &
          \multicolumn{2}{c}{\textbf{\,\,\,\,Problematic Apps\,\,\,\,}} \\ \cline{4-11}
         &
           &
           &
          ST &
          IC &
          SL &
          UI &
          TK &
          IS &
          \% &
          \#/\# \\ \Xhline{1pt}
        \rowcolor{lightgray!50} \multicolumn{11}{c}{\textbf{Unsystematic Prompt/Query Construction}} \\ \hline
        Unclear context in prompt &
          §\ref{subsec:Improper context in prompt} &
          LLM agent &
          \multicolumn{1}{c}{} &
          \checkmark &
           &
           &
           &
           &
          77\% &
          77/100 \\ \hline
        Imprecise knowledge retrieval &
          §\ref{subsec:Improper similarity search} &
          vector database &
          \multicolumn{1}{c}{} &
          \checkmark &
           &
           &
          \checkmark&
           &
         45\% &
          45/100 \\ \hline
         \rowcolor{lightgray!50} \multicolumn{11}{c}{\textbf{Misunderstanding of Interface Specifications}} \\ \hline
        Missing LLM input format validation &
          §\ref{subsec:Missing LLM input format validation} &
          LLM agent &
          \checkmark &
          \checkmark &
           &
           &
           &
           &
          83\% &
          83/100 \\ \hline
        Incompatible LLM output format &
          §\ref{subsec:Incompatible LLM output format} &
          LLM agent &
          \checkmark &
          \checkmark &
           &
          \checkmark &
           &
           &
          26\% &
          26/100 \\ \hline
        Unnecessary LLM output &
          §\ref{subsec:Unnecessary LLM output} &
          LLM agent &
           &
           &
          \checkmark &
          \checkmark &
          \checkmark &
           &
          36\% &
          36/100 \\ \hline
        Exceeding LLM context limit &
          §\ref{subsec:Exceeding LLM context limit} &
          LLM agent &
          \checkmark &
          \checkmark &
           &
           &
           &
           &
          85\% &
          85/100 \\ \hline
        Knowledge misalignment &
          §\ref{subsec:Tokenization misalignment} &
          vector database &
          \multicolumn{1}{c}{} &
          \checkmark &
           &
           &
           &
           &
          22\% &
          22/100 \\ \hline
        Conflicting knowledge entries &
          §\ref{subsec:Conflict data entries} &
          vector database &
          \checkmark &
          \checkmark &
           &
           &
           &
          \checkmark &
          \,\,\,8\% &
          \,\,\,8/100 \\ \hline
        Improper text embedding &
          §\ref{subsec:Improper embedding} &
          vector database &
          \checkmark &
          \checkmark &
           &
           &
          \checkmark &
           &
          15\% &
          15/100 \\ \hline
        Low-frequency interactivity &
          §\ref{subsec:Low-Frequency interactivity} &
          Software components &
          \checkmark &
           &
           &
           &
           &
           &
          90\% &
          11/\,\,\,13 \\ \hline
        \rowcolor{lightgray!50}\multicolumn{11}{c}{\textbf{Unaware of Software Context}} \\ \hline
        Lacking restrictions in prompt &
          §\ref{subsec:Lacking restrictions in prompt} &
          LLM agent &
          \multicolumn{1}{c}{} &
          \checkmark &
           &
           &
           &
           &
          14\% &
          14/100 \\ \hline
        Insufficient history management &
          §\ref{subsec:Insufficient history management} &
          LLM agent &
          \multicolumn{1}{c}{} &
          \checkmark &
           &
           &
           &
           &
          44\% &
          44/100 \\ \hline
        Absence of final output &
          §\ref{subsec:Without final output} &
          Software components &
          \multicolumn{1}{c}{} &
           &
           &
          \checkmark &
           &
           &
          88\% &
          21/\,\,\,24 \\ \hline
        Sketchy error handling &
          §\ref{subsec:Improper error handling} &
          Software components &
          \checkmark &
           &
           &
          &
           &
           &
          10\% &
          10/100 \\ \hline
        \rowcolor{lightgray!50}\multicolumn{11}{c}{\textbf{Lacking System Management}} \\ \hline
        Privacy violation &
          §\ref{subsec:Privacy Violation} &
          \,\,\,\,Software components\,\,\,\, &
          \multicolumn{1}{c}{} &
           &
           &
           &
           &
          \checkmark &
          53\% &
          10/\,\,\,19 \\ \hline
        Resource contention &
          §\ref{subsec:Resource contention} &
          system &
          \checkmark &
           &
          \checkmark  &
           &
           &
           &
          92\% &
          12/\,\,\,13 \\ \hline
        Inefficient memory management &
          §\ref{subsec:Inefficient memory management} &
          system &
          \checkmark &
           &
          \checkmark &
           &
           &
           &
          23\% &
          23/100 \\ \hline
        Out-of-sync LLM downstream tasks \,\,\,\,\,\,\,\,\,\,\, &
          §\ref{subsec:Mismatched speed between LLM and downstream tasks} &
          system &
          \checkmark &
           &
          \checkmark &
           &
           &
           &
          77\% &
          10/\,\,\,13 \\ \hline
        \Xhline{1pt}
        \multicolumn{9}{c}{Total number of benchmark applications with \emph{more than three} types of defects} &
          77\% &
          77/100 \\ \Xhline{1pt}
        \end{tabular}
        \begin{tablenotes}
            \item * In the \emph{Impact} column, from left to right, ST refer to fail-stops, IC refer to incorrectness, SL refer to slower execution, UI refer to unfriendly user interface, TK refer to more tokens, and IS refer to insecure.
            \item * In the \emph{Problematic Apps} column, the denominator refers to applications that should make efforts to prevent such defect, and the numerator refers to applications that actually contain such defect (details in the corresponding sections).
        \end{tablenotes}
    \end{threeparttable}}
\end{table*}

%% file: sections/5-lesson.tex
\label{sec:lesson}

We provide a systematic guideline for developers to reveal and resolve the defects in LLM-enabled software. 
Our future work will explore automated solutions.

\medskip
\noindent\emph{Lesson 1: Specifying Component Interfaces}

The components of LLM-enabled software interact with each other, necessitating clear specifications of their required inputs and expected behaviors. When initializing the vector database, chunking and embedding algorithms should be carefully designed to support various input formats and contents (§\ref{subsec:Tokenization misalignment},\ref{Improper embedding}). For data pipeline and timing behaviors, the software should incorporate queue and heart-beat techniques to {tolerate} various processing {speeds} (§\ref{subsec:Low-Frequency interactivity},\ref{Mismatched speed between LLM and downstream tasks}). A modular design is also recommended to facilitate LLM and RAG updates, software iteration, and system maintenance.

\medskip
\noindent\emph{Lesson 2: Pre-Processing Text Inputs}

The quality of prompts significantly impacts LLM agent behaviors and software quality. Several defects could be alleviated through input pre-processing (§\ref{subsec:Insufficient history management},\ref{Missing LLM input format validation},\ref{Unnecessary LLM output},\ref{Exceeding LLM context limit}). 
The input prompt should include clear instructions and response templates to avoid unnecessary outputs, as well as a short summary of past conversations for in-context {responses}.
When extracting text strings from various file formats, input validation and conversion are required to fit LLM capability.

\medskip
\noindent\emph{Lesson 3: Post-Processing to Refine Outputs}
 
LLMs may produce responses that are incorrect or unfit for downstream tasks (§\ref{subsec:Incompatible LLM output format}). Therefore, the LLM agent should design task-specific solutions to re-order and re-structure the LLM outputs. Sometimes, incorrect LLM outputs could be fixed using software context~\cite{wan2023smartgear}. 
Knowledge entries retrieved from vector databases also require relevance validation (through similarity scores or NLP techniques) to filter out unrequired ones (§\ref{subsec:Improper similarity search}).

\medskip
\noindent\emph{Lesson 4: Thinking in Larger Scale}

LLM and RAG algorithms are executed within the software context and system environment. When developing a component, the developers should have the entire software system in mind (§\ref{subsec:Without final output}).
To mitigate cascade effects of unexpected component outputs, robust exception handling and recovery mechanisms are required, including service up/down-grades and LLM regeneration with failure information (§\ref{subsec:Improper error handling}). Efficient resource management is also required to improve efficiency and prevent hangs (§\ref{subsec:Resource contention},\ref{Inefficient memory management}).

\medskip
\noindent\emph{Lesson 5: Thorough Alpha/Beta Testing}

Given the huge input space and unpredictable behavior of LLM-enabled software, testing is essential to reveal software defects, especially for defects in §\ref{subsec:Improper context in prompt},\ref{Lacking restrictions in prompt},\ref{Conflict data entries}. During alpha testing, developers should design test cases that cover various surface-level patterns and semantics of LLM prompts and knowledge entries. Performance and security testing are also required to detect efficiency and security problems.

As 320 of 546 defects of our empirical study were reported by the end-users, we also recommend beta testing with participants from various backgrounds.

{
\medskip
\noindent\emph{Lesson 6: Automatic Detection and Repairing}

The behavior of LLM-enabled applications is influenced by both control-flow and data-flow. Defects related to control-flow can be detected and patched through static analysis, as illustrated in Figure~\ref{fig:solution for improper text embedding} and \ref{fig:solution for inefficient memory management}. The ones related to data-flow should be addressed at run-time. 
For instance, \emph{lacking restrictions in prompts} and \emph{sketchy error handling} can be mitigated using advanced prompting techniques and feedback mechanisms to guide LLMs in self-correcting their outputs; \emph{exceeding LLM context limit} can be resolved by direct intervention in the inference process; and \emph{incompatible LLM output format} can be localized by tracing data-flow.

}

%% file: sections/8-discussion.tex
\textbf{Internal threats to validity.} The test inputs may not represent the actual workload. The issues collected and confirmed in our study may not encompass all software defects. Our manual defect identification may contain biases.

\textbf{External threats to validity.} Our study is limited to LLMs for general language and code tasks, excluding multi-modal and fined-tuned LLMs. It only covers a subset of vector databases and frameworks. 
We only study open-source projects on GitHub, excluding closed-source commercial ones. Our benchmark may not represent all real-world applications.

%% file: sections/9-related_work.tex
\subsection{LLM and RAG}

LLMs have revolutionized various AI research fields and enabled advanced applications~\cite{brown2020language, achiam2023gpt}, including human-like conversations~\cite{kamalloo-etal-2023-evaluating, lin-chen-2023-llm}, mathematical reasoning~\cite{imani-etal-2023-mathprompter}, planning~\cite{zhao2024large}, control~\cite{song2023llm}, and software engineering~\cite{lemieux2023codamosa, hou2023large, feng2024prompting, fan2023automated, cong2023Invalidator, xia2024fuzz4all, DBLP:journals/pacmpl/ZhangCGLPSV24}. 
Recently, researchers designed fine-tuning~\cite{luo2024wizardcoder}, prompt engineering~\cite{ding2024cycle} and ensemble~\cite{hong2023metagpt, yang2024swe} techniques to improve LLM capability on specific tasks. However, LLMs still face trustworthiness and efficiency problems~\cite{zhang2023hallucination, wang2023decodingtrust, liu2024exploring}.
RAG techniques~\cite{gao2023retrieval, zhao2024retrievalaugmented} were then proposed to enhance LLMs with external knowledge, in question-answering~\cite{lewis2020retrieval}, planning~\cite{lee2024planrag}, coding~\cite{parvez-etal-2021-retrieval-augmented, zhou2023docprompting, zhang-etal-2023-repocoder, wang2023rap}, and other tasks.
These works focus on LLM and RAG algorithm designs, instead of how to integrate them into software systems.

One recent work~\cite{liu2023demystifying} explored remote code execution vulnerabilities of LLM frameworks. It only studied the security problems. In contrast, our work studies a broader scope of engineering challenges and integration failures in real-world applications that incorporate LLM and RAG techniques.

\subsection{AI-enabled software}
Prior works have studied the development and maintenance of AI-enabled software throughout its lifecycle. 

One line of works improved the quality of neural networks, through testing~\cite{pei2017deepxplore, ma2018deepgauge, tian2018deeptest, kim2019guiding}, monitoring~\cite{xiao2021self, stocco2022thirdeye, huang2023patchcensor} and repairing~\cite{zhang2019apricot, usman2021nn, zhang2021autotrainer, qi2023archrepair}. 
Some works studied deep learning frameworks~\cite{pham2019cradle, xie2022docter, liu2023generation, wei2024demystifying} and compilers~\cite{liu2023nnsmith, wang2023gencog}. These works focus on constructing AI models rather than using them.

Another line of works explored the usage of AI in real-world applications. Several works studied the development challenges~\cite{zhang2019empirical, alshangiti2019developing} and deployment problems~\cite{chen2020comprehensive, humbatova2020taxonomy} of AI-enabled software, including mobile applications~\cite{xu2019first, chen2021empirical}, cluster infrastructures~\cite{zhang2020empirical}, and general systems~\cite{guo2019empirical}.
Other works studied the integration of cloud AI services~\cite{wan2021machine, chen2022efficient, xie2022cost, wan2023smartgear}.
All of these studies focus on traditional AI designed for specific tasks, which typically have categorical or structural outputs. They do not address the unique challenges of general-purpose LLMs.

%% file: sections/10-conclusion.tex
LLMs with RAG support have been widely integrated into real-world applications. This paper conducts the first comprehensive study of integration challenges of LLM and RAG techniques. We have investigated 100 open-source LLM-enabled software and manually studied 3,000 issue reports, finding that integration defects are widespread and severe. We summarize 18 defect patterns that cause functionality, efficiency, and security problems. We also construct a defect library \defectlib, offering guidance to resolve integration failures. We believe that this work will aid the development of LLM-enabled software and motivate future research.

%% file: main.bbl
\begin{thebibliography}{100}
\providecommand{\url}[1]{#1}
\csname url@samestyle\endcsname
\providecommand{\newblock}{\relax}
\providecommand{\bibinfo}[2]{#2}
\providecommand{\BIBentrySTDinterwordspacing}{\spaceskip=0pt\relax}
\providecommand{\BIBentryALTinterwordstretchfactor}{4}
\providecommand{\BIBentryALTinterwordspacing}{\spaceskip=\fontdimen2\font plus
\BIBentryALTinterwordstretchfactor\fontdimen3\font minus \fontdimen4\font\relax}
\providecommand{\BIBforeignlanguage}[2]{{%
\expandafter\ifx\csname l@#1\endcsname\relax
\typeout{** WARNING: IEEEtran.bst: No hyphenation pattern has been}%
\typeout{** loaded for the language `#1'. Using the pattern for}%
\typeout{** the default language instead.}%
\else
\language=\csname l@#1\endcsname
\fi
#2}}
\providecommand{\BIBdecl}{\relax}
\BIBdecl

\bibitem{Hydrangea}
\BIBentryALTinterwordspacing
Y.~Shao, ``Hydrangea: A defect library,'' 2024, accessed: 6 February 2025. [Online]. Available: \url{https://github.com/ycshao12/Hydrangea}
\BIBentrySTDinterwordspacing

\bibitem{openai2024}
{OpenAI}, ``{OpenAI: Advancing Digital Intelligence},'' \url{https://openai.com/}, 2024, accessed: 2024-06-20.

\bibitem{LlamaChat}
A.~Rozanski, ``Llamachat: A chat application using llama framework,'' \url{https://github.com/alexrozanski/LlamaChat}, 2024.

\bibitem{langchain2024}
\BIBentryALTinterwordspacing
L.~Team, \emph{LangChain: A Library for Language Data Processing}, 1st~ed.\hskip 1em plus 0.5em minus 0.4em\relax Global: LangChain Press, 2024. [Online]. Available: \url{https://www.langchain.com/}
\BIBentrySTDinterwordspacing

\bibitem{zirnstein2023extended}
B.~Zirnstein, ``Extended context for instructgpt with llamaindex,'' 2023.

\bibitem{mongodb2024}
{MongoDB, Inc.}, ``{MongoDB: The Developer Data Platform},'' \url{https://www.mongodb.com/}, 2024, accessed: 2024-06-20.

\bibitem{chroma2024}
{Chroma}, ``{Chroma: The AI-native open-source embedding database},'' \url{https://www.trychroma.com/}, 2024, accessed: 2024-06-20.

\bibitem{faiss2023}
LangChain, ``Faiss,'' \url{https://python.langchain.com/v0.2/docs/integrations/vectorstores/faiss/}, 2023, accessed: 2024-07-03.

\bibitem{wang2023decodingtrust}
\BIBentryALTinterwordspacing
B.~Wang, W.~Chen, H.~Pei \emph{et~al.}, ``Decodingtrust: A comprehensive assessment of trustworthiness in {GPT} models,'' in \emph{Thirty-seventh Conference on Neural Information Processing Systems Datasets and Benchmarks Track}, 2023. [Online]. Available: \url{https://openreview.net/forum?id=kaHpo8OZw2}
\BIBentrySTDinterwordspacing

\bibitem{sun2024trustllm}
L.~Sun, Y.~Huang, H.~Wang, S.~Wu, Q.~Zhang, C.~Gao, Y.~Huang, W.~Lyu, Y.~Zhang, X.~Li \emph{et~al.}, ``Trustllm: Trustworthiness in large language models,'' \emph{arXiv preprint arXiv:2401.05561}, 2024.

\bibitem{yao2024survey}
Y.~Yao, J.~Duan, K.~Xu, Y.~Cai, Z.~Sun, and Y.~Zhang, ``A survey on large language model (llm) security and privacy: The good, the bad, and the ugly,'' \emph{High-Confidence Computing}, p. 100211, 2024.

\bibitem{wan2021machine}
C.~Wan, S.~Liu, H.~Hoffmann, M.~Maire, and S.~Lu, ``Are machine learning cloud apis used correctly?'' in \emph{ICSE}, 2021.

\bibitem{chen2022efficient}
L.~Chen, M.~Zaharia, and J.~Zou, ``Efficient online ml api selection for multi-label classification tasks,'' in \emph{ICML}.\hskip 1em plus 0.5em minus 0.4em\relax PMLR, 2022, pp. 3716--3746.

\bibitem{xie2022cost}
S.~Xie, Y.~Xue, Y.~Zhu, and Z.~Wang, ``Cost effective mlaas federation: A combinatorial reinforcement learning approach,'' in \emph{IEEE INFOCOM 2022-IEEE Conference on Computer Communications}.\hskip 1em plus 0.5em minus 0.4em\relax IEEE, 2022, pp. 2078--2087.

\bibitem{wan2023smartgear}
\BIBentryALTinterwordspacing
C.~Wan, Y.~Liu, K.~Du, H.~Hoffmann, J.~Jiang, M.~Maire, and S.~Lu, ``Run-time prevention of software integration failures of machine learning apis,'' \emph{Proc. ACM Program. Lang.}, vol.~7, no. OOPSLA2, oct 2023. [Online]. Available: \url{https://doi.org/10.1145/3622806}
\BIBentrySTDinterwordspacing

\bibitem{luo2024wizardcoder}
Z.~Luo, C.~Xu \emph{et~al.}, ``Wizardcoder: Empowering code large language models with evol-instruct,'' in \emph{ICLR}, 2024.

\bibitem{ding2024cycle}
Y.~Ding, M.~J. Min, G.~Kaiser, and B.~Ray, ``Cycle: Learning to self-refine the code generation,'' \emph{Proceedings of the ACM on Programming Languages}, vol.~8, no. OOPSLA1, pp. 392--418, 2024.

\bibitem{gao2023retrieval}
Y.~Gao, Y.~Xiong, X.~Gao, K.~Jia, J.~Pan, Y.~Bi, Y.~Dai, J.~Sun, and H.~Wang, ``Retrieval-augmented generation for large language models: A survey,'' \emph{arXiv preprint arXiv:2312.10997}, 2023.

\bibitem{zhao2024retrievalaugmented}
P.~Zhao, H.~Zhang, Q.~Yu, Z.~Wang, Y.~Geng, F.~Fu, L.~Yang, W.~Zhang, and B.~Cui, ``Retrieval-augmented generation for ai-generated content: A survey,'' 2024.

\bibitem{bubeck2023sparks}
S.~Bubeck, V.~Chandrasekaran, R.~Eldan, J.~Gehrke, E.~Horvitz, E.~Kamar, P.~Lee, Y.~T. Lee, Y.~Li, S.~Lundberg \emph{et~al.}, ``Sparks of artificial general intelligence: Early experiments with gpt-4,'' \emph{arXiv preprint arXiv:2303.12712}, 2023.

\bibitem{shaunwei2024realchar}
S.~Wei, ``{RealChar: A Realistic Character Simulation Toolkit},'' \url{https://github.com/Shaunwei/RealChar}, 2024, accessed: 2024-06-20.

\bibitem{raj2019fundamentals}
P.~Raj, A.~Raman, and D.~Kakadia, \emph{Fundamentals of Software Integration}.\hskip 1em plus 0.5em minus 0.4em\relax CRC Press, 2019.

\bibitem{winters2020software}
T.~Winters, T.~Manshreck, and H.~Wright, \emph{Software Engineering at Google: Lessons Learned from Programming Over Time}.\hskip 1em plus 0.5em minus 0.4em\relax O'Reilly Media, 2020.

\bibitem{llama_meta}
\BIBentryALTinterwordspacing
Meta, ``Llama,'' 2024, democratizing access through an open platform featuring AI models, tools, and resources to give people the power to shape the next wave of innovation. Licensed for both research and commercial use. [Online]. Available: \url{https://llama.meta.com/}
\BIBentrySTDinterwordspacing

\bibitem{zhang2023hallucination}
Y.~Zhang, Y.~Li, L.~Cui, D.~Cai, Liu \emph{et~al.}, ``Siren's song in the ai ocean: A survey on hallucination in large language models,'' \emph{arXiv preprint arXiv:2309.01219}, 2023.

\bibitem{ChatIQ2024}
Y.~Saka, ``Chatiq: Versatile slack bot with gpt and weaviate-powered long-term memory,'' \url{https://github.com/yujiosaka/ChatIQ}, 2024, accessed: 2024-06-23.

\bibitem{white2024prompt}
\BIBentryALTinterwordspacing
J.~White, Q.~Fu, S.~Hays, M.~Sandborn, C.~Olea, H.~Gilbert, A.~Elnashar, J.~Spencer-Smith, and D.~C. Schmidt, ``A prompt pattern catalog to enhance prompt engineering with chatgpt,'' \emph{arXiv preprint arXiv:2302.11382}, 2024. [Online]. Available: \url{https://arxiv.org/abs/2302.11382}
\BIBentrySTDinterwordspacing

\bibitem{mayooear_gpt4}
Mayooear, ``Pdf-chatbot,'' \url{https://github.com/mayooear/gpt4-pdf-chatbot-langchain}, 2024.

\bibitem{nakajima_babyagi}
Y.~Nakajima, ``Babyagi,'' \url{https://github.com/yoheinakajima/babyagi}, 2024.

\bibitem{FOLLGAD_Godmode-GPT}
FOLLGAD, ``Godmode-gpt,'' \url{https://github.com/FOLLGAD/Godmode-GPT}, 2023.

\bibitem{mayooear_gpt4_pdf_chatbot_langchain}
mayooear, ``Gpt-4 pdf chatbot using langchain,'' \url{https://github.com/mayooear/gpt4-pdf-chatbot-langchain}, 2024, gitHub.

\bibitem{h2oai_h2ogpt}
H2O.ai, ``h2ogpt,'' \url{https://github.com/h2oai/h2ogpt}, 2024.

\bibitem{privateGPT2024}
Z.~AI, ``privategpt: Secure and private interactions with large language models,'' \url{https://github.com/zylon-ai/private-gpt}, 2024, accessed: 2024-06-23.

\bibitem{vicuna7b}
LMSYS, ``Vicuna-7b model card,'' \url{https://huggingface.co/lmsys/vicuna-7b-v1.5}, 2024, accessed: 2024-07-04.

\bibitem{LLMChat}
LLMChat, ``Llmchat: A full-stack webui implementation of large language model,'' \url{https://github.com/c0sogi/LLMChat}, 2024.

\bibitem{mattzcarey_code_review_gpt}
M.~Carey, ``code-review,'' \url{https://github.com/mattzcarey/code-review-gpt}, 2024.

\bibitem{anthropic2024claude}
{Anthropic}, ``{Claude: An AI Assistant by Anthropic},'' \url{https://www.anthropic.com/claude}, 2024, accessed: 2024-06-20.

\bibitem{quivr2024}
QuivrHQ, ``Quivr: Open-source rag framework for building genai second brains,'' \url{https://github.com/QuivrHQ/quivr}, 2024, accessed: 2024-06-23.

\bibitem{RobbyChatbot}
Y.~Ba, ``Robby-chatbot,'' \url{https://github.com/yvann-ba/Robby-chatbot}, 2024.

\bibitem{yang2018hotpotqa}
Z.~Yang, P.~Qi, S.~Zhang \emph{et~al.}, ``{HotpotQA}: A dataset for diverse, explainable multi-hop question answering,'' in \emph{EMNLP}, 2018.

\bibitem{barnett2024seven}
S.~Barnett, S.~Kurniawan, S.~Thudumu, Z.~Brannelly, and M.~Abdelrazek, ``Seven failure points when engineering a retrieval augmented generation system,'' in \emph{CAIN}, 2024, pp. 194--199.

\bibitem{autogpt}
S.~Gravitas, ``Autogpt,'' \url{https://github.com/Significant-Gravitas/AutoGPT}, 2024, accessed: 2024-06-22.

\bibitem{fastgpt}
L.~Team, ``Fastgpt: A knowledge-based platform built on the llm,'' \url{https://github.com/labring/FastGPT}, 2024, accessed: 2024-07-04.

\bibitem{LangChainChatGLMWebui}
X-D-Lab, ``Langchain chatglm webui: A web user interface for langchain chatglm,'' \url{https://github.com/X-D-Lab/LangChain-ChatGLM-Webui}, 2024.

\bibitem{anything-llm}
M.~Labs, ``Anything-llm,'' \url{https://github.com/Mintplex-Labs/anything-llm}, 2024, accessed: 2024-06-26.

\bibitem{LangchainChatchat}
C.~Team, ``Chatchat: Local knowledge-based qa with langchain and chatglm,'' \url{https://github.com/chatchat-space/Langchain-Chatchat}, 2024.

\bibitem{ChatDocs}
Marella, ``Chatdocs: Chat with your documents offline using ai,'' \url{https://github.com/marella/chatdocs}, 2024.

\bibitem{LocalAGI}
EmbraceAGI, ``Localagi: Local deployment of agi tools,'' \url{https://github.com/EmbraceAGI/LocalAGI}, 2024.

\bibitem{xue2024db}
S.~Xue, C.~Jiang, W.~Shi, F.~Cheng, K.~Chen, H.~Yang, Z.~Zhang, J.~He, H.~Zhang, G.~Wei, W.~Zhao, F.~Zhou, D.~Qi, H.~Yi, S.~Liu, and F.~Chen, ``Db-gpt: Empowering database interactions with private large language models,'' \emph{arXiv preprint arXiv:2312.17449}, 2024.

\bibitem{100birdspecies}
P.~Gopiński, ``100 bird species dataset,'' \url{https://www.kaggle.com/datasets/gpiosenka/100-bird-species}, 2024, accessed: 2024-07-05.

\bibitem{urllib3}
urllib3 Contributors, ``urllib3: Http library with thread-safe connection pooling, file post, and more,'' \url{https://pypi.org/project/urllib3/}, 2024, accessed: 2024-07-05.

\bibitem{jiang2023promptpackerdeceivingllms}
\BIBentryALTinterwordspacing
S.~Jiang, X.~Chen, and R.~Tang, ``Prompt packer: Deceiving llms through compositional instruction with hidden attacks,'' 2023. [Online]. Available: \url{https://arxiv.org/abs/2310.10077}
\BIBentrySTDinterwordspacing

\bibitem{langchain_chatbot}
H.~Corporation, ``Langchain chatbot,'' \url{https://github.com/Haste171/langchain-chatbot}, 2023, accessed: 2024-07-05.

\bibitem{bhawal_resume_dataset}
\BIBentryALTinterwordspacing
S.~Bhawal, ``Resume dataset,'' 2023, accessed: 2024-07-29. [Online]. Available: \url{https://www.kaggle.com/datasets/snehaanbhawal/resume-dataset}
\BIBentrySTDinterwordspacing

\bibitem{brown2020language}
T.~Brown, B.~Mann, N.~Ryder, M.~Subbiah, J.~D. Kaplan, P.~Dhariwal, A.~Neelakantan, P.~Shyam, G.~Sastry, A.~Askell \emph{et~al.}, ``Language models are few-shot learners,'' \emph{Advances in neural information processing systems}, vol.~33, pp. 1877--1901, 2020.

\bibitem{achiam2023gpt}
J.~Achiam, S.~Adler, S.~Agarwal, L.~Ahmad, I.~Akkaya, F.~L. Aleman, D.~Almeida, J.~Altenschmidt, S.~Altman, S.~Anadkat \emph{et~al.}, ``Gpt-4 technical report,'' \emph{arXiv preprint arXiv:2303.08774}, 2023.

\bibitem{kamalloo-etal-2023-evaluating}
E.~Kamalloo, N.~Dziri, C.~Clarke, and D.~Rafiei, ``Evaluating open-domain question answering in the era of large language models,'' in \emph{ACL}, A.~Rogers, J.~Boyd-Graber, and N.~Okazaki, Eds., 2023.

\bibitem{lin-chen-2023-llm}
Y.-T. Lin and Y.-N. Chen, ``{LLM}-eval: Unified multi-dimensional automatic evaluation for open-domain conversations with large language models,'' in \emph{NLP4ConvAI}, Y.-N. Chen and A.~Rastogi, Eds., 2023.

\bibitem{imani-etal-2023-mathprompter}
S.~Imani, L.~Du, and H.~Shrivastava, ``{M}ath{P}rompter: Mathematical reasoning using large language models,'' in \emph{Proceedings of the 61st Annual Meeting of the Association for Computational Linguistics (Volume 5: Industry Track)}, S.~Sitaram, B.~Beigman~Klebanov, and J.~D. Williams, Eds.\hskip 1em plus 0.5em minus 0.4em\relax Toronto, Canada: Association for Computational Linguistics, Jul. 2023.

\bibitem{zhao2024large}
Z.~Zhao, W.~S. Lee, and D.~Hsu, ``Large language models as commonsense knowledge for large-scale task planning,'' \emph{Advances in Neural Information Processing Systems}, vol.~36, 2024.

\bibitem{song2023llm}
C.~H. Song, J.~Wu, C.~Washington, B.~M. Sadler, W.-L. Chao, and Y.~Su, ``Llm-planner: Few-shot grounded planning for embodied agents with large language models,'' in \emph{ICCV}, 2023.

\bibitem{lemieux2023codamosa}
C.~Lemieux, J.~P. Inala, S.~K. Lahiri, and S.~Sen, ``Codamosa: Escaping coverage plateaus in test generation with pre-trained large language models,'' in \emph{ICSE}.\hskip 1em plus 0.5em minus 0.4em\relax IEEE, 2023, pp. 919--931.

\bibitem{hou2023large}
X.~Hou, Y.~Zhao, Y.~Liu, Z.~Yang, K.~Wang, L.~Li, X.~Luo, D.~Lo, J.~Grundy, and H.~Wang, ``Large language models for software engineering: A systematic literature review,'' \emph{arXiv preprint arXiv:2308.10620}, 2023.

\bibitem{feng2024prompting}
S.~Feng and C.~Chen, ``Prompting is all you need: Automated android bug replay with large language models,'' in \emph{ICSE}, 2024, pp. 1--13.

\bibitem{fan2023automated}
Z.~Fan, X.~Gao, M.~Mirchev, A.~Roychoudhury, and S.~H. Tan, ``Automated repair of programs from large language models,'' in \emph{ICSE}, 2023, pp. 1469--1481.

\bibitem{cong2023Invalidator}
T.~Le-Cong, D.-M. Luong, X.~B.~D. Le, D.~Lo, N.-H. Tran, B.~Quang-Huy, and Q.-T. Huynh, ``Invalidator: Automated patch correctness assessment via semantic and syntactic reasoning,'' \emph{TSE}, 2023.

\bibitem{xia2024fuzz4all}
C.~S. Xia, M.~Paltenghi, J.~Le~Tian, M.~Pradel, and L.~Zhang, ``Fuzz4all: Universal fuzzing with large language models,'' \emph{Proc. IEEE/ACM ICSE}, 2024.

\bibitem{DBLP:journals/pacmpl/ZhangCGLPSV24}
J.~Zhang, J.~P. Cambronero, S.~Gulwani, V.~Le, R.~Piskac, G.~Soares, and G.~Verbruggen, ``Pydex: Repairing bugs in introductory python assignments using llms,'' \emph{OOPSLA1}, 2024.

\bibitem{hong2023metagpt}
S.~Hong, X.~Zheng, J.~Chen, Y.~Cheng, J.~Wang, C.~Zhang, Z.~Wang, S.~K.~S. Yau, Z.~Lin, L.~Zhou \emph{et~al.}, ``Metagpt: Meta programming for multi-agent collaborative framework,'' \emph{arXiv preprint arXiv:2308.00352}, 2023.

\bibitem{yang2024swe}
J.~Yang, C.~E. Jimenez, A.~Wettig, K.~Lieret, S.~Yao, K.~Narasimhan, and O.~Press, ``Swe-agent: Agent-computer interfaces enable automated software engineering,'' \emph{arXiv preprint arXiv:2405.15793}, 2024.

\bibitem{liu2024exploring}
F.~Liu, Y.~Liu, L.~Shi, H.~Huang, R.~Wang, Z.~Yang, and L.~Zhang, ``Exploring and evaluating hallucinations in llm-powered code generation,'' \emph{arXiv preprint arXiv:2404.00971}, 2024.

\bibitem{lewis2020retrieval}
P.~Lewis, E.~Perez, A.~Piktus \emph{et~al.}, ``Retrieval-augmented generation for knowledge-intensive nlp tasks,'' \emph{Advances in Neural Information Processing Systems}, vol.~33, pp. 9459--9474, 2020.

\bibitem{lee2024planrag}
M.~Lee, S.~An, and M.-S. Kim, ``Planrag: A plan-then-retrieval augmented generation for generative large language models as decision makers,'' in \emph{NAACL}, 2024, pp. 6537--6555.

\bibitem{parvez-etal-2021-retrieval-augmented}
M.~R. Parvez, W.~Ahmad, S.~Chakraborty, B.~Ray, and K.-W. Chang, ``Retrieval augmented code generation and summarization,'' in \emph{EMNLP}, 2021, pp. 2719--2734.

\bibitem{zhou2023docprompting}
S.~Zhou, U.~Alon, F.~F. Xu, Z.~Jiang, and G.~Neubig, ``Docprompting: Generating code by retrieving the docs,'' in \emph{ICLR}, 2023.

\bibitem{zhang-etal-2023-repocoder}
F.~Zhang, B.~Chen, Y.~Zhang, J.~Keung, J.~Liu, D.~Zan, Y.~Mao, J.-G. Lou, and W.~Chen, ``{R}epo{C}oder: Repository-level code completion through iterative retrieval and generation,'' in \emph{EMNLP}, H.~Bouamor, J.~Pino, and K.~Bali, Eds., 2023.

\bibitem{wang2023rap}
W.~Wang, Y.~Wang, S.~Joty, and S.~C. Hoi, ``Rap-gen: Retrieval-augmented patch generation with codet5 for automatic program repair,'' in \emph{ESEC/FSE}, 2023, pp. 146--158.

\bibitem{liu2023demystifying}
T.~Liu, Z.~Deng, G.~Meng, Y.~Li, and K.~Chen, ``Demystifying rce vulnerabilities in llm-integrated apps,'' \emph{arXiv preprint arXiv:2309.02926}, 2023.

\bibitem{pei2017deepxplore}
K.~Pei, Y.~Cao, J.~Yang, and S.~Jana, ``Deepxplore: Automated whitebox testing of deep learning systems,'' in \emph{proceedings of the 26th Symposium on Operating Systems Principles}, 2017, pp. 1--18.

\bibitem{ma2018deepgauge}
L.~Ma, F.~Juefei-Xu, F.~Zhang, J.~Sun, M.~Xue, B.~Li, C.~Chen, T.~Su, L.~Li, Y.~Liu \emph{et~al.}, ``Deepgauge: Multi-granularity testing criteria for deep learning systems,'' in \emph{ASE}, 2018, pp. 120--131.

\bibitem{tian2018deeptest}
Y.~Tian, K.~Pei, S.~Jana, and B.~Ray, ``Deeptest: Automated testing of deep-neural-network-driven autonomous cars,'' in \emph{ICSE}, 2018.

\bibitem{kim2019guiding}
J.~Kim, R.~Feldt, and S.~Yoo, ``Guiding deep learning system testing using surprise adequacy,'' in \emph{ICSE}.\hskip 1em plus 0.5em minus 0.4em\relax IEEE, 2019, pp. 1039--1049.

\bibitem{xiao2021self}
Y.~Xiao, I.~Beschastnikh, D.~S. Rosenblum, C.~Sun, S.~Elbaum, Y.~Lin, and J.~S. Dong, ``Self-checking deep neural networks in deployment,'' in \emph{ICSE}, 2021.

\bibitem{stocco2022thirdeye}
A.~Stocco, P.~J. Nunes, M.~d'Amorim, and P.~Tonella, ``Thirdeye: Attention maps for safe autonomous driving systems,'' in \emph{ASE}, 2022, pp. 1--12.

\bibitem{huang2023patchcensor}
Y.~Huang, L.~Ma, and Y.~Li, ``Patchcensor: Patch robustness certification for transformers via exhaustive testing,'' \emph{ACM Transactions on Software Engineering and Methodology}, vol.~32, no.~6, pp. 1--34, 2023.

\bibitem{zhang2019apricot}
H.~Zhang and W.~Chan, ``Apricot: A weight-adaptation approach to fixing deep learning models,'' in \emph{ASE}.\hskip 1em plus 0.5em minus 0.4em\relax IEEE, 2019, pp. 376--387.

\bibitem{usman2021nn}
M.~Usman, D.~Gopinath, Y.~Sun, Y.~Noller, and C.~S. P{\u{a}}s{\u{a}}reanu, ``Nn repair: Constraint-based repair of neural network classifiers,'' in \emph{CAV}, 2021, pp. 3--25.

\bibitem{zhang2021autotrainer}
X.~Zhang, J.~Zhai, S.~Ma, and C.~Shen, ``Autotrainer: An automatic dnn training problem detection and repair system,'' in \emph{ICSE}, 2021.

\bibitem{qi2023archrepair}
H.~Qi, Z.~Wang, Q.~Guo, J.~Chen, F.~Juefei-Xu, F.~Zhang, L.~Ma, and J.~Zhao, ``Archrepair: Block-level architecture-oriented repairing for deep neural networks,'' \emph{ACM TSE}, vol.~32, no.~5, pp. 1--31, 2023.

\bibitem{pham2019cradle}
H.~V. Pham, T.~Lutellier, W.~Qi, and L.~Tan, ``Cradle: cross-backend validation to detect and localize bugs in deep learning libraries,'' in \emph{ICSE}, 2019, pp. 1027--1038.

\bibitem{xie2022docter}
D.~Xie, Y.~Li, M.~Kim, H.~V. Pham, L.~Tan, X.~Zhang, and M.~W. Godfrey, ``Docter: Documentation-guided fuzzing for testing deep learning api functions,'' in \emph{ISSTA}, 2022, pp. 176--188.

\bibitem{liu2023generation}
J.~Liu, Y.~Huang, Z.~Wang, L.~Ma, C.~Fang, M.~Gu, X.~Zhang, and Z.~Chen, ``Generation-based differential fuzzing for deep learning libraries,'' \emph{ACM TOSEM}, vol.~33, no.~2, pp. 1--28, 2023.

\bibitem{wei2024demystifying}
M.~Wei, N.~S. Harzevili, Y.~Huang, J.~Yang, J.~Wang, and S.~Wang, ``Demystifying and detecting misuses of deep learning apis,'' in \emph{ICSE}, 2024, pp. 1--12.

\bibitem{liu2023nnsmith}
J.~Liu, J.~Lin, F.~Ruffy, C.~Tan, J.~Li, A.~Panda, and L.~Zhang, ``Nnsmith: Generating diverse and valid test cases for deep learning compilers,'' in \emph{ASPLOS}, 2023, pp. 530--543.

\bibitem{wang2023gencog}
Z.~Wang, P.~Nie, X.~Miao, Y.~Chen, C.~Wan, L.~Bu, and J.~Zhao, ``Gencog: A dsl-based approach to generating computation graphs for tvm testing,'' in \emph{ISSTA}, 2023, pp. 904--916.

\bibitem{zhang2019empirical}
T.~Zhang, C.~Gao, L.~Ma, M.~Lyu, and M.~Kim, ``An empirical study of common challenges in developing deep learning applications,'' in \emph{ISSRE}.\hskip 1em plus 0.5em minus 0.4em\relax IEEE, 2019, pp. 104--115.

\bibitem{alshangiti2019developing}
M.~Alshangiti, H.~Sapkota, P.~K. Murukannaiah, X.~Liu, and Q.~Yu, ``Why is developing machine learning applications challenging? a study on stack overflow posts,'' in \emph{ESEM}.\hskip 1em plus 0.5em minus 0.4em\relax IEEE, 2019, pp. 1--11.

\bibitem{chen2020comprehensive}
Z.~Chen, Y.~Cao, Y.~Liu, H.~Wang, T.~Xie, and X.~Liu, ``A comprehensive study on challenges in deploying deep learning based software,'' in \emph{ESEC/FSE}, 2020, pp. 750--762.

\bibitem{humbatova2020taxonomy}
N.~Humbatova, G.~Jahangirova, G.~Bavota, V.~Riccio, A.~Stocco, and P.~Tonella, ``Taxonomy of real faults in deep learning systems,'' in \emph{ICSE}, 2020, pp. 1110--1121.

\bibitem{xu2019first}
M.~Xu, J.~Liu, Y.~Liu, F.~X. Lin, Y.~Liu, and X.~Liu, ``A first look at deep learning apps on smartphones,'' in \emph{WWW}, 2019, pp. 2125--2136.

\bibitem{chen2021empirical}
Z.~Chen, H.~Yao, Y.~Lou, Y.~Cao, Y.~Liu, H.~Wang, and X.~Liu, ``An empirical study on deployment faults of deep learning based mobile applications,'' in \emph{ICSE}.\hskip 1em plus 0.5em minus 0.4em\relax IEEE, 2021, pp. 674--685.

\bibitem{zhang2020empirical}
R.~Zhang, W.~Xiao, H.~Zhang, Y.~Liu, H.~Lin, and M.~Yang, ``An empirical study on program failures of deep learning jobs,'' in \emph{ICSE}, 2020, pp. 1159--1170.

\bibitem{guo2019empirical}
Q.~Guo, S.~Chen, X.~Xie, L.~Ma, Q.~Hu, H.~Liu, Y.~Liu, J.~Zhao, and X.~Li, ``An empirical study towards characterizing deep learning development and deployment across different frameworks and platforms,'' in \emph{ASE}.\hskip 1em plus 0.5em minus 0.4em\relax IEEE, 2019, pp. 810--822.

\end{thebibliography}
